\documentclass[openacc]{rstransa}

\titlehead{Review}

\usepackage{comment}
\usepackage{siunitx}
\usepackage{tabularx}
\usepackage{ragged2e}
\usepackage{booktabs}
\usepackage{amsmath}
\newcolumntype{L}[1]{>{\raggedright\arraybackslash}p{#1}}
\newcolumntype{C}[1]{>{\centering\arraybackslash}p{#1}}
\newcolumntype{P}[1]{>{\RaggedRight\arraybackslash}p{#1}}
\usepackage{url}
\DeclareSIUnit{\rpm}{rpm}
\DeclareSIUnit{\rad}{rad}
\usepackage{dcolumn}
\usepackage{bm}
\usepackage{soul}
\usepackage{ulem} 
\usepackage{lipsum}  
\usepackage[numbers]{natbib} 

\definecolor{autor1color}{RGB}{220,20,60}   
\definecolor{autor2color}{RGB}{0,102,204}   
\definecolor{autor3color}{RGB}{34,139,34}   
\definecolor{autor4color}{RGB}{255,140,0}   
\definecolor{autor5color}{RGB}{128,0,128}   

\begin{document}
\sloppy
\title{Kinetics of coagulation phenomena from a granular matter perspective}
\author{
 Gustavo Castillo$^{1}$ and Nicol\'as Mujica$^{2}$}

\address{$^{1}$Instituto de Ciencias de la Ingeniería, Universidad de O'Higgins,  Rancagua, Chile\\
$^{2}$Departamento de Física, Facultad de Ciencias Físicas y Matemáticas, Universidad de Chile, Santiago, Chile}

\subject{Fluid mechanics, Mechanics, Statistical physics}


\corres{Gustavo Castillo\\
\email{gustavo.castillo@uoh.cl}}

\begin{abstract}

Aggregation processes play a central role in systems ranging from
aerosol coagulation and cloud formation to dust growth in
protoplanetary disks and granular materials. These processes are
traditionally described by Smoluchowski’s coagulation equation, which
provides a mean-field account of growth through binary collisions.
However, incorporation of granular physics—dissipative
interactions, spatial heterogeneity, and force transmission through
contact networks—reveals important limitations of this framework.

In this review, we show how such effects lead to the breakdown of
mean-field assumptions and motivate a view of aggregation as a
multi-scale process shaped by the interplay between interactions,
structure, and collective dynamics. Phenomena such as segregation,
jamming, and clogging further highlight the role of mechanical
constraints and spatial organization in limiting or redirecting
growth. By integrating insights from granular physics, aerosol
science, and astrophysics, we outline a unified perspective on
coagulation in non-equilibrium particulate systems.

This paper is part of the thematic issue ''Sand, silos and asteroids: clustering challenges in granular materials research''.

\end{abstract}


\begin{fmtext}
\section{Introduction}

\subsection{Beyond Smoluchowski: From Classical Coagulation to Granular Aggregation}

\par The theoretical description of particle coagulation dates back to the pioneering work of Marian Smoluchowski in 1916, who formulated the kinetic equation governing the time evolution of cluster populations undergoing binary aggregation~\cite{smoluchowski1916drei}. His mean-field framework provided a statistical description of how particles collide and merge, establishing what is now known as the Smoluchowski coagulation equation. Since then, coagulation phenomena have been recognized as central to a wide range of physical systems. 
\end{fmtext}
\maketitle

\par In astrophysics, aggregation processes are fundamental to the early stages of planet formation through dust coagulation in protoplanetary disks~\cite{guttler2010outcome-dfe, Drazkowska2014, Birnstiel2016, tominaga2025dust}. In atmospheric science, particle coagulation influences the dynamics and size distribution of aerosols and cloud droplets~\cite{Fuchs, friedlander2000smoke, chen2017agglomeration-591, vemury1997coagulation-03a, PhysRevFluids.7.064308}. Industrial processes such as aerosol synthesis, powder processing, and colloidal stabilization are also strongly governed by coagulation kinetics~\cite{buesser2012design, buesser2009coagulation, camelot1999bipolar, adachi1995dynamic, fortuny2004coagulation}. Over the decades, Smoluchowski’s original mean-field framework has been extended and generalized to incorporate effects such as size-dependent kernels, spatial correlations, fractal aggregate structures, and additional interaction forces, giving rise to a broad family of kinetic models used to describe aggregation phenomena across disciplines~\cite{drake1972general, leyvraz2003scaling, vicsek1992fractal, PhysRevA.40.3836, PhysRevLett.51.1123, osinsky2022exact, szala2023modified}. 

Blum~\cite{Blum2006} provided a comprehensive and influential review of dust agglomeration, systematically establishing the kinetic description of the aggregation processes based on the Smoluchowski equation, with particular emphasis on astrophysical dust in gaseous environments. That work clarified the role of aggregation kernels, fractal aggregate structures, and sticking mechanisms, and remains a cornerstone reference for studies of dust growth in space, aerosols, and related particulate systems. However, its primary focus was on mean-field descriptions of well-mixed ensembles, where aggregation kinetics can be reduced to effective binary collision rates and their consequences for mass distributions and growth laws.
In the two decades since that review, substantial progress has been made in granular matter physics, nonequilibrium statistical mechanics, and particle-resolved numerical methods, motivating a complementary perspective on coagulation phenomena. In particular, aggregation is now increasingly understood as a clustering process in dissipative particulate systems, where spatial inhomogeneities, dissipation, long-range interactions, and feedback between structure and kinetics play a central role~\cite{PhysRevFluids.9.034304, brilliantov2010kinetic, Carrillo2021}.\\
\indent In parallel with developments in aggregation theory, the field of granular physics experienced rapid growth during the 1990s and early 2000s, providing new conceptual tools to describe collections of dissipative particles. Early theoretical advances emphasized the role of inelastic collisions in dilute granular gases and led to the introduction of the notion of \textit{granular temperature}, a statistical measure of velocity fluctuations that allowed kinetic-theory approaches to be extended to dissipative systems~\cite{Goldhirsch2008}. At the same time, studies of dense granular matter revealed collective phenomena such as the jamming–unjamming transition and the emergence of heterogeneous force networks that transmit stress through contact chains~\cite{Cates1998, Majmudar2007, Biroli2007}. These developments substantially improved the understanding of how microscopic collision rules give rise to macroscopic structure and dynamics in particulate systems. Many of these concepts have begun to intersect with coagulation research. In particular, ideas from granular kinetic theory, clustering instabilities, and contact network formation have provided new perspectives for interpreting aggregation processes and the evolution of cluster structures in systems where collisions are dissipative and interactions extend beyond simple mean-field assumptions~\cite{dufty2000statistical, brilliantov2010kinetic}.\\
Several recent developments make this an opportune time to revisit coagulation phenomena from a granular matter perspective. Granular kinetic theory has matured substantially, providing improved frameworks for describing dissipative collisions, clustering, and collective particle dynamics in a wide range of densities~\cite{brilliantov2010kinetic,poschel2005computational}. At the same time, major observational breakthroughs have transformed our empirical understanding of aggregation processes. High-resolution observations from facilities such as Atacama Large Millimeter/submillimeter Array have revealed unprecedented details of dust evolution in protoplanetary disks, while experiments conducted in microgravity platforms including the International Space Station have enabled controlled studies of particle aggregation under conditions difficult to reproduce on Earth~\cite{alma20152014, andrews2018disk, testi2014dust, blum2000growth, brisset2017nanorocks, wurm2021understanding-87f,Schubert_2024}. Advances in experimental techniques have also played a key role: high-speed imaging, particle-tracking methods, and surface and bulk material characterizations, such as atomic force microscopy, are now widely accessible, enabling more precise characterization of collision dynamics, surface roughness, and forces, as well as aggregate growth~\cite{Blum2008, Royer2009}. Finally, rapid progress in computational capabilities—particularly the widespread use of GPUs and the integration of machine learning techniques—has dramatically expanded the scale and duration of particle-resolved simulations, allowing larger and longer discrete element method (DEM) studies and opening new possibilities for data-driven analysis of coagulation processes~\cite{Lu2021, wang2025machine,ZHOU2023118969}.

\begin{table}[t!]
\centering
\caption{Conceptual evolution of coagulation theory from classical mean-field descriptions to a granular matter perspective.}
\label{tab:then-now}
\begin{tabular}{P{0.28\textwidth} P{0.3\textwidth} P{0.3\textwidth}}
\hline
\textbf{Aspect} & \textbf{Then (classical view, e.g.\ Blum 2006)} & \textbf{Now (granular matter perspective)} \\
\hline
Primary framework &
Smoluchowski equation with effective coagulation kernels &
Coagulation as a clustering process in dissipative particulate systems \\

Role of spatial structure &
Assumed homogeneous and well-mixed; spatial effects treated as corrections &
Spatial inhomogeneities emerge dynamically and feed back into kinetics \\

Interpretation of kernels &
Prescribed collision rates encoding microscopic physics &
Emergent, history-dependent effective interactions shaped by clustering \\

Dissipation &
Mainly controls sticking thresholds and restructuring &
Central control parameter governing irreversibility, clustering and arrest \\

Aggregate structure &
Fractal dimension commonly used as a static descriptor (though dynamic treatments exist~\cite{1993AA280617O}) &
Structure evolves dynamically and couples back to collision rates \\

Long-range interactions (e.g. electric charges) &
Modify kernels via enhancement or suppression factors &
Lead to frustration, kinetic arrest and non-universal long-time behaviour \\

Growth regimes &
Hit-and-stick, compaction and fragmentation treated as distinct regimes &
Continuous crossover between regimes governed by dissipation and interactions \\

Universality and scaling &
Emphasis on self-similar solutions and universality classes &
Universality limited by correlations, long-range forces and dissipation \\

Numerical approaches &
Mean-field solvers and early Monte Carlo methods &
Particle-based simulations and hybrid kinetic--granular models \\

Scope of applicability &
Primarily dust in gaseous environments (astrophysical focus) &
Unified description across aerosols, granular media and astrophysical systems \\
\hline
\end{tabular}
\end{table}

Table~\ref{tab:then-now} summarizes the conceptual shift from classical mean-field descriptions of coagulation, as reviewed by Blum~\cite{Blum2006}, to a granular matter perspective that emphasizes dissipation, clustering, and emergent kinetics.\\

\subsection{Scope and Special Issue Context}

\par In this review, we revisit classical coagulation phenomena through the conceptual framework of granular physics. Our goal is not to provide an exhaustive survey of the vast aggregation literature, but rather to offer a focused synthesis that highlights how ideas from granular matter can enrich the understanding of coagulating systems. In particular, we emphasize kinetic aspects of the process, including collision dynamics, energy dissipation, clustering, and the mechanisms governing particle growth. By examining aggregation through the lens of dissipative particle interactions and collective granular dynamics, we aim to connect traditional mean-field descriptions with insights emerging from granular kinetic theory and particle-scale observations. This perspective allows us to reinterpret familiar coagulation processes in terms of collision-driven dynamics and energy transport, identifying common principles that bridge fields such as aerosol science, planetary formation, and particulate processing. The review therefore seeks to provide a strategic framework that brings together key concepts from both communities while highlighting open questions and opportunities for future research. We focus on aggregation in systems whose dynamics remain at least mildly
compatible with a kinetic description---granular gases, dilute
suspensions, and astrophysical dust populations---where adaptations of the
Smoluchowski framework remain meaningful. Aggregation in dense, continuum
granular flows, which is strongly out of equilibrium and generally
requires problem-specific treatments, lies beyond the scope of the present
review.\\
\indent Granular materials provide a unifying physical framework linking systems that at first glance appear very different, from laboratory sand experiments to industrial silos and planetary bodies. On Earth, controlled experiments with sand and other granular media have long served as analog systems to investigate particle-scale interactions, collision dynamics, and collective behaviour under well-defined conditions. In confined geometries such as silos and hoppers, granular materials exhibit striking phenomena including force-chain networks, clogging, and intermittent flow, revealing how stresses propagate through disordered particle assemblies~\cite{Janda2009, majmudar2005contact-9b7, LopezRodriguez2019}. At much larger scales, similar principles appear to govern the structure of small planetary bodies. Many asteroids are now understood to possess “rubble-pile” structures composed of loosely bound aggregates and regolith layers, where gravity replaces confinement but particle contacts and collective mechanics remain central~\cite{richardson2002gravitational, walsh2018rubble}. Insights from sample-return missions and in situ measurements increasingly support this granular interpretation~\cite{yada2022preliminary, sugita2019geomorphology}. Across these contexts, the common thread is that macroscopic behaviour emerges from the interactions of many discrete grains, making granular physics a powerful lens through which to interpret both terrestrial and planetary aggregation processes.\\
\indent The remainder of this review is organized as follows. Section 2 introduces the theoretical foundations that underpin coagulation and granular dynamics, outlining the key concepts and models that form the basis for the discussion that follows. Section 3 examines collision dynamics, focusing on particle–particle interactions, restitution, and the mechanisms through which dissipative collisions influence aggregation outcomes. Section 4 explores the kinetics of aggregation, the structure of aggregates, including fractal growth, force-chain organization and the emergence of heterogeneous internal structures. Section 5 highlights applications across different physical systems, illustrating how these ideas manifest in contexts ranging from laboratory granular media to astrophysical environments. Section 6 provides a perspective on developments over the past two decades, revisiting key advances and shifts in understanding. Finally, Section 7 discusses open challenges and future directions, with Section 8 summarizing the main conclusions of the review.

\section{Theoretical Framework: Unifying Coagulation and Granular Physics}
Theoretical descriptions of aggregation processes have traditionally been formulated within kinetic frameworks that treat particle growth as a sequence of binary collisions in a statistically homogeneous ensemble. In this approach, the evolution of particle populations is commonly described by rate equations derived from Smoluchowski’s coagulation theory, which provide a mean-field representation of aggregation dynamics. While this framework has been widely applied in fields ranging from aerosol physics to astrophysical dust growth, it relies on assumptions such as spatial homogeneity and weak correlations. These assumptions become less evident in dissipative particulate systems, where clustering, collective motion and spatial structure can emerge. In this context, it is useful to revisit the classical Smoluchowski framework and examine its interpretation within a broader granular matter perspective.
\subsection{Smoluchowski Framework Revisited}
\begin{figure}
    \centering
    \includegraphics[width=\linewidth]{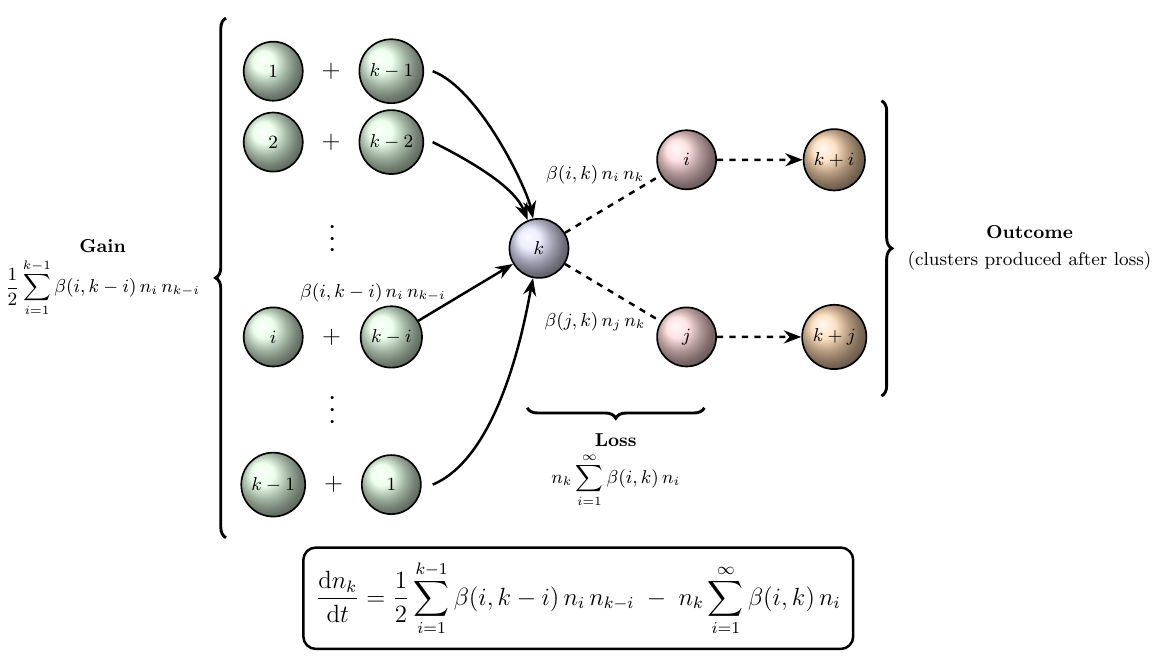}
    \caption{Conceptual diagram of particle aggregation kinetics described by the Smoluchowski coagulation equation. The schematic illustrates how binary collisions between clusters lead to the formation (gain term) and removal (loss term) of aggregates of a given size, which together determine the temporal evolution of the cluster population. The aggregation kernel $\beta_{ij}$ represents the collision rate between clusters of size $i$ and $j$, incorporating the physical processes that govern particle encounters and sticking, such as relative motion, interaction forces, and dissipative collision dynamics.}
    \label{fig:Eq:smolu}
\end{figure}

The kinetic description of aggregation processes is commonly formulated in terms of the Smoluchowski growth equation, which describes the temporal evolution of the concentration of clusters of different sizes. Let $n_i(t)$ denote the concentration of clusters containing $i$ primary particles at time $t$, i.e. their number per unit volume (or per
unit area in two dimensions). The rate of change of $n_i$ results from two competing mechanisms: the formation of clusters of size $i$ through collisions between smaller aggregates, and the disappearance of clusters of size $i$ when they collide with other particles to form larger clusters. These processes are represented by gain and loss terms in the kinetic equation,
\begin{equation}
\frac{dn_i}{dt}
=
\frac{1}{2}\sum_{j=1}^{i-1} \beta_{j,i-j}\, n_j n_{i-j}
-
n_i \sum_{j=1}^{\infty} \beta_{ij}\, n_j ,
\label{eq:smoluchowski}
\end{equation}
where $\beta_{ij}$ denotes the collision kernel describing the rate at which clusters of sizes $i$ and $j$ interact ($\beta_{ij}$ has units of volume (or area) per unit time). The first term on the right-hand side represents the gain of clusters of size $i$ resulting from collisions between clusters of sizes $j$ and $i-j$, while the factor $1/2$ avoids double counting of symmetric collision pairs. The second term accounts for the loss of clusters of size $i$ when they collide with clusters of any other size $j$ to form larger aggregates. A schematic representation of these collision processes is shown in Fig.~\ref{fig:Eq:smolu}. In the absence of external sources or sinks, this equation conserves the total mass of the system, so that the quantity
\begin{equation}
M_1 = \sum_{i=1}^{\infty} i\, n_i(t)
\end{equation}
remains constant in time, even though the total number of clusters decreases as aggregation proceeds.

The collision kernel $\beta_{ij}$ encapsulates the physical processes that govern the interaction between clusters of sizes $i$ and $j$. In general, it represents the rate at which such particles collide and therefore depends on the mechanisms that generate relative motion as well as on the geometrical properties of the interacting particles. For instance, in ballistic aggregation, the kernel can be expressed in terms of the collision cross section $\sigma_{ij}$, the relative velocity $v_{ij}$, and a sticking probability or collision efficiency $\alpha_{ij}$,
\begin{equation}
\beta_{ij} = \alpha_{ij}\,\sigma_{ij}\,v_{ij}.
\label{eq:kernel_general}
\end{equation}
In this form, the kernel highlights the physical origin of coagulation rates: particles collide more frequently if they move faster, present larger effective cross sections, or stick efficiently upon contact. The dependence of $\sigma_{ij}$ and $v_{ij}$ on particle size and environmental conditions determines the functional form of the kernel and ultimately controls the kinetics of aggregation.

Different physical mechanisms therefore give rise to different forms of the collision kernel, as illustrated by the representative examples summarized in Table~\ref{tab:kernels}. In dilute systems, where particle motion is dominated by thermal fluctuations, aggregation is typically driven by Brownian motion, leading to kernels that depend on the diffusivities and sizes of the interacting particles. In systems where particles move along nearly straight trajectories, collisions may instead be described by ballistic kernels in which the collision rate is proportional to the geometric cross section and the relative velocity of the particles. Turbulent flows provide another important mechanism for generating relative velocities, particularly in atmospheric and astrophysical environments, where turbulent shear and inertial effects can significantly enhance collision rates. 

Additional interactions may further modify the effective collision rate. For instance, electrostatic forces between charged particles can either enhance or suppress coagulation depending on the sign and magnitude of the charges. In such cases, the Brownian collision kernel may be corrected by factors accounting for Coulomb interactions between particles, leading to modified kernels that capture the influence of long-range forces on aggregation dynamics. These examples illustrate that the collision kernel serves as the central link between microscopic interaction physics and the macroscopic evolution of the particle size distribution described by the Smoluchowski equation.

\begin{table}[t]
\centering
\caption{Representative collision kernels commonly used in coagulation theory. 
The kernel $\beta_{ij}$ represents the collision rate between clusters of sizes 
$i$ and $j$, and depends on the physical mechanism generating relative motion 
and the effective interaction cross section between particles. For gravitational focusing, the escape velocity corresponds to ${v_\text{esc}=\sqrt{2G(M_i+M_j)/(r_i+r_j)}}$.}
\label{tab:kernels}
\begin{tabular}{p{2cm} p{4.6cm} p{3cm} p{3cm}}
\toprule
\textbf{Mechanism} & \textbf{Kernel expression} & \textbf{Physical parameters} & \textbf{Typical references} \\
\midrule

Constant kernel &
$\beta_{ij} = K$ &
Independent of particle size; often used as a theoretical benchmark &
Smoluchowski (1916)~\cite{smoluchowski1916drei}; Leyvraz (2003)~\cite{leyvraz2003scaling}\\

\addlinespace

3D Brownian diffusion &
$\displaystyle
\beta_{ij} = 4\pi (D_i + D_j)(r_i + r_j)
$ &
Diffusion coefficients $D_i$, particle radii $r_i$; thermal motion &
Smoluchowski (1918)~\cite{Smoluchowski1918}; Friedlander (2000)~\cite{friedlander2000smoke}\\

\addlinespace

Ballistic (geometric) &
$\displaystyle
\beta_{ij} = \pi (r_i + r_j)^2\, v_{ij}
$ &
Geometric cross section and relative velocity &
Safronov (1972)~\cite{Safronov1972}; Blum (2006)~\cite{Blum2006}\\

\addlinespace

Turbulent collision &
$\displaystyle
\beta_{ij} \sim \pi (r_i+r_j)^2 \, \Delta v_{ij}^{\rm turb}
$ &
Turbulent velocity fluctuations; particle inertia &
Saffman \& Turner (1956)~\cite{SaffmanTurner1956}; Falkovich et al. (2002)~\cite{Falkovich2002}\\

\addlinespace

Brownian + Coulomb interaction &
$\displaystyle
\beta_{ij} = \beta_{ij}^{\rm B} \, W_{ij}
$ &
Brownian diffusion kernel $\beta^{\rm B}$ modified by electrostatic factor $W_{ij}$ &
Ivlev et al. (2002)~\cite{ivlev2002coagulation-51a}; Matthews \& Hyde (2009)~\cite{Matthews2009}\\

\addlinespace

Gravitational focusing &
$\displaystyle
\beta_{ij} = \pi (r_i+r_j)^2 v_{ij}
\left(1 + \frac{v_{\rm esc}^2}{v_{ij}^2}\right)
$ &
Gravitational attraction enhances collision cross section &
Safronov (1972)~\cite{Safronov1972}; Wetherill (1989)~\cite{wetherill1989accumulation-7f6}\\

\bottomrule
\end{tabular}
\end{table}

The Smoluchowski equation admits a number of characteristic solution regimes that provide important insight into the kinetics of aggregation. In general, however, the equation does not admit analytical solutions for physically realistic kernels. Exact solutions exist only for a small number of idealized cases, such as constant, additive, or multiplicative kernels, which serve mainly as theoretical benchmarks~\cite{leyvraz2003scaling}. As a result, the evolution of particle populations in most practical situations must be investigated using numerical approaches. Common strategies include direct numerical integration of the discretized Smoluchowski equation~\cite{Birnstiel2010, Okuzumi2009} and stochastic particle-based methods, particularly Monte Carlo algorithms, which simulate the sequence of binary collision events in a statistically representative ensemble~\cite{Kruis2000}. 

In its continuum formulation, the Smoluchowski equation takes the form of a nonlinear integro--differential equation for the particle concentration distribution $n(m,t)$, which further complicates analytical treatment. Despite these challenges, several general classes of solutions have been identified that characterize aggregation dynamics. In simplified monodisperse descriptions, where the particle population is approximated by clusters of nearly equal size, the mean cluster mass can exhibit simple growth laws determined by the scaling properties of the collision kernel~\cite{dongen1988scaling-b3e, ernst1986fractals-a2a}. Depending on the underlying collision mechanism, the mean mass may grow algebraically or exponentially in time; for example, Brownian-driven aggregation typically leads to parabolic growth of the characteristic cluster size, while kernels associated with constant relative velocities can produce exponential growth. In more realistic polydisperse systems, the particle size distribution often evolves toward a self-similar form in which the entire distribution can be expressed as a universal function of the rescaled mass $i/\langle i(t) \rangle$, where $\langle i(t) \rangle$ denotes the mean cluster mass (ensemble average
over the cluster population at time $t$) This dynamic scaling behaviour reflects the emergence of statistical self-similarity in the aggregation process. A further important regime occurs when the kernel grows sufficiently rapidly with particle size, leading to the phenomenon of \textit{gelation}, in which a finite fraction of the total mass condenses into a macroscopic cluster within finite time~\cite{Schubert_2024, lushnikov2006gelation-506}. In such cases the growth becomes dominated by runaway aggregation, marking a qualitative departure from the self-similar scaling regime.

\subsection{Granular Physics Essentials}
Granular materials constitute a broad class of particulate systems whose dynamics are governed by interactions that differ fundamentally from those encountered in molecular fluids. A useful classification is given by Andreotti et al. \cite{andreotti2013granular}, where particles of size larger than $\SI{100}{\micro\meter}$ are considered as granular media, for which in many situations dissipative collisions and frictional contacts are dominant. In this classification, particles of sizes between 1 and $\SI{100}{\micro\meter}$ are considered as powders and below $\SI{1}{\micro\meter}$ we refer to colloids. For the former, van der Waals forces, capillary bridge forces and air drag are more relevant than the collisional interactions. For the latter, the thermal nature of the systems is recovered. In particular, for granular media collisions between grains are typically inelastic, leading to a systematic dissipation of kinetic energy through a variety of microscopic mechanisms \cite{BrilliantovPoschel2004, Johnson_1985}. As a result, the collective behaviour of granular systems reflects a delicate interplay between contact mechanics, energy dissipation, and large-scale particle motion \cite{Jaeger1996,Aranson2006}. Understanding these ingredients is essential when attempting to connect granular dynamics with aggregation phenomena. In this section we therefore review several key physical concepts that underpin the mechanics of granular interactions. 

We begin by examining the physics of particle contacts, where classical elastic contact theory provides the starting point for describing interactions between deformable grains. Models such as the Hertzian contact law capture the elastic response of colliding particles \cite{Hertz1882}, while adhesive extensions including the Johnson--Kendall--Roberts (JKR) framework account for the role of surface forces in cohesive particulate systems \cite{JKR1971}. Beyond purely elastic behaviour, a range of dissipative processes may occur during collisions, including plastic deformation, viscoelastic losses, and microscopic surface rearrangements at the contact interface \cite{BrilliantovPoschel2004}. These mechanisms determine the degree of energy dissipation during particle encounters and strongly influence the subsequent evolution of the system.

Building on this microscopic picture, we then consider the statistical description of many-particle granular systems through the framework of granular kinetic theory. In analogy with molecular gases, this approach introduces the concept of a granular temperature to quantify velocity fluctuations, while explicitly accounting for the irreversible loss of energy during collisions \cite{BrilliantovPoschel2004,Goldhirsch2003}. Granular systems often require external energy injection to sustain a steady state, and under many conditions they exhibit striking collective phenomena such as clustering instabilities and spatially heterogeneous particle distributions \cite{GoldhirschZanetti1993}. These features highlight an important contrast with the homogeneous mixing assumptions that typically underlie classical coagulation models. In the final part of this section we synthesize these perspectives by examining how concepts from granular physics can inform the kinetic description of aggregation processes. In particular, we discuss how dissipation, correlations, and collective motion may lead to departures from mean-field coagulation theory, and we outline how modern computational approaches---including discrete element simulations and particle-based stochastic methods---provide powerful tools for exploring this interface between granular dynamics and coagulation kinetics \cite{CundallStrack1979,Kruis2000}.

\subsubsection{Inelastic Collisions and Energy Dissipation}
A central quantity used to characterize the inelastic nature of particle collisions in granular systems is the coefficient of restitution. This dimensionless parameter measures the fraction of relative normal velocity that is preserved after a binary collision. If two particles approach each other with a normal relative velocity $v_n$ before impact and separate with velocity $v_n'$ after the collision, the coefficient of restitution is defined as
\begin{equation}
e = -\frac{v_n'}{v_n}.
\end{equation}
The value of $e$ therefore quantifies the degree of energy dissipation occurring during the contact. In the limit $e=1$, collisions are perfectly elastic and kinetic energy is conserved, as in ideal molecular gases. At the opposite extreme, $e=0$ corresponds to perfectly inelastic collisions in which particles stick together after impact, converting all normal relative motion into deformation or internal energy.

In many real granular materials the coefficient of restitution is not a constant, but depends on the impact velocity, the mechanical properties of the interacting particles and surface roughness, with values typically
ranging between $e \approx 0.5$ and $0.95$~\cite{BrilliantovPoschel2004, schwager2008coefficient-d8c,Montaine2011}. Experiments and theoretical models show that, for viscoelastic spheres, energy dissipation during contact arises from internal friction and time-dependent deformation, leading to a restitution coefficient that decreases with increasing impact velocity. Within viscoelastic contact theory, this behaviour can be captured by expanding the restitution coefficient as a function of the normal impact velocity $v_n$,
\begin{equation}\label{eq: rest}
e(v_n) \simeq 1 - C\, v_n^{1/5} + \mathcal{O}(v_n^{2/5}),
\end{equation}
where the constant $C$ depends on the elastic moduli and dissipative material parameters of the particles. This velocity dependence plays an important role in granular kinetic theory, since it directly affects the rate of energy dissipation in a system of colliding grains and influences the evolution of the granular temperature. As a result, models that incorporate velocity-dependent restitution provide a more realistic description of granular gases and dense particulate flows than formulations based on a constant restitution coefficient.
It is important to note that the expansion in Eq.~\ref{eq: rest} neglects cohesive interactions and is therefore only valid for sufficiently large impact velocities where adhesion can be ignored. When cohesive forces become relevant, the coefficient of restitution exhibits a non-monotonic dependence on impact velocity, reaching a maximum at intermediate velocities and tending to zero both at low velocities (due to sticking) and at high velocities (due to enhanced dissipative mechanisms such as plastic deformation)~\cite{nietiadi2020bouncing}. Such effects are particularly important in fine or highly cohesive particulate systems.

The mechanics of particle contacts provides the microscopic foundation for understanding collisions and aggregation in particulate systems. In the simplest case of purely elastic deformation without adhesion, the interaction between two spherical particles is described by the classical Hertz contact theory, which relates the normal force $F$ to the overlap $\delta$ between the particles through the nonlinear relation $F \propto \delta^{3/2}$ \cite{Hertz1882}. While this description is appropriate for relatively large grains or collisions occurring at sufficiently high impact velocities, surface forces become increasingly important as particle sizes decrease. For small particles, adhesive interactions arising from van der Waals forces can significantly modify the contact mechanics and collision outcomes. Two widely used models extend the Hertzian framework to include adhesion. The Johnson--Kendall--Roberts (JKR) model describes contacts where adhesive forces act primarily within the contact area and is typically relevant for relatively soft materials or larger particles, such as micron-sized dust grains in astrophysical or aerosol environments \cite{JKR1971}. In contrast, the Derjaguin--Muller--Toporov (DMT) model assumes that adhesion acts mainly outside the contact region and is more appropriate for stiffer materials or smaller particles with weak surface deformation \cite{DMT1975}. These contact theories play a central role in determining collision outcomes in particulate systems, particularly in environments such as protoplanetary disks or dusty plasmas where particle sizes span regimes in which adhesive forces strongly influence aggregation dynamics~\cite{Dominik1997}.

Even when the contact mechanics of colliding particles can be described by elastic models, real collisions between granular particles are rarely perfectly conservative. A variety of dissipative mechanisms may operate during the contact process, converting kinetic energy into internal deformation, heat, or microscopic surface restructuring. One important source of dissipation arises from viscoelastic losses within the material, where time-dependent deformation leads to hysteresis in the loading and unloading phases of the contact \cite{BrilliantovPoschel1996}. In many materials, particularly metals or brittle grains at sufficiently high impact velocities, plastic deformation may also occur, permanently altering the shape of the contact region and further increasing energy dissipation. In addition to bulk deformation, surface processes can contribute significantly to energy loss. Microscopic roughness, surface asperities, and the rearrangement or breakage of surface bonds may dissipate energy even when the overall deformation remains elastic~\cite{greenwood1966contact-73f, Israelachvili2011}. These mechanisms are particularly important when viewed from the perspective of collision outcomes. In many kinetic descriptions of aggregation, the effect of microscopic dissipation is captured through a sticking probability $\alpha$, which represents the likelihood that a collision results in the formation of a stable contact rather than rebound. The various dissipative processes discussed above---including viscoelastic losses, plastic deformation, and surface rearrangements at the contact interface---act to remove kinetic energy from the relative motion of the colliding particles, thereby increasing the probability that the particles remain bound after impact. Conversely, when dissipation is weak, a significant fraction of collisions may result in rebound, leading to a reduced sticking efficiency. The sticking probability therefore provides a convenient bridge between microscopic collision physics and macroscopic aggregation kinetics, since it directly enters the definition of the collision kernel (see Eq.~\ref{eq:kernel_general}). In this way, the detailed contact mechanics and dissipative processes occurring during collisions ultimately control the effective aggregation rates that appear in kinetic models of coagulation~\cite{Blum2008}.

These considerations emphasize that the outcome of individual collisions
plays a central role in determining the effective aggregation kinetics.
When large ensembles of particles interact through such dissipative
collisions, their collective dynamics can be described within the
framework of granular kinetic theory, which we briefly review next.

\subsubsection{Granular Kinetic Theory}

When large ensembles of particles interact through dissipative collisions, their collective dynamics can be described within the framework of granular kinetic theory. While this approach shares many conceptual similarities with the kinetic theory of classical dilute gases, the presence of inelastic collisions introduces fundamental differences in the statistical behaviour of the system. In particular, the loss of kinetic energy during collisions leads to a gradual decay of particle velocity fluctuations in an isolated system, a process commonly referred to as \textit{granular cooling}~\cite{goldhirsch1993clustering-dc7}. 

In analogy with molecular systems, the velocity fluctuations of particles are characterized by the \textit{granular temperature}, defined as
\begin{equation}
T_g = \frac{m}{d}\langle \mathbf{v'}^2 \rangle ,
\end{equation}
where $m$ is the particle mass, $d$ the number of spatial dimensions, and $\langle \mathbf{v'}^2 \rangle$ the mean square fluctuating velocity relative to the average flow. Despite its name, the granular temperature is not a thermodynamic temperature: it does not correspond to thermal equilibrium with a heat bath, but instead provides a measure of the kinetic energy associated with the random motion of grains. Because collisions are dissipative, a freely evolving granular gas continually loses kinetic energy and cools down in the absence of external energy input. As a consequence, granular systems generally do not reach equilibrium states analogous to those of molecular gases. Classical results of equilibrium statistical mechanics, such as the equipartition theorem, therefore do not hold in general, and mixtures of particles with different masses or sizes may exhibit unequal kinetic temperatures.

A kinetic description of dilute granular gases can nevertheless be constructed by extending the Boltzmann framework to account for inelastic collisions. In this formulation, the evolution of the particle velocity distribution is described by a Boltzmann--Enskog equation for the probability density function of particle velocities, derived under the usual assumption of molecular chaos \cite{goldshtein1995mechanics-4b0}. From this kinetic equation, hydrodynamic equations for density, momentum, and granular temperature can be obtained through suitable moment expansions. 

Beyond the dilute regime, alternative statistical descriptions have been proposed for dense granular matter~\cite{baule2018edwards, Makse2004}. In particular, Edwards and collaborators introduced a statistical mechanics framework in which the ensemble of mechanically stable configurations is characterized by the system volume rather than energy, with the volume fraction playing a role analogous to the Hamiltonian in equilibrium statistical mechanics. Within this approach, the concept of \textit{compactivity} emerges as a thermodynamic-like variable conjugate to volume \cite{EdwardsGrinev1999}. Although this framework provides a compelling theoretical perspective on dense granular systems, the experimental determination of compactivity remains challenging, and its precise physical interpretation continues to be debated \cite{Gramlich2023}.

Even in dilute granular gases, the dissipative nature of collisions can lead to striking collective behaviour absent in molecular systems. One of the most prominent examples is the emergence of clustering instabilities in freely cooling granular gases. Small fluctuations in the local particle density or velocity field can grow over time because regions with slightly higher density experience more frequent collisions and therefore dissipate kinetic energy more rapidly. As particles in these regions lose kinetic energy, their velocities decrease and they take longer to escape, while particles streaming in from
hotter neighbouring regions enter faster than they leave; this asymmetry,
driven purely by inelastic collisions rather than by any attractive
interaction, produces a net accumulation of particles in the cooler region. This positive feedback mechanism amplifies the initial density fluctuations and eventually leads to the formation of clusters or dense regions embedded within a dilute background \cite{GoldhirschZanetti1993}. The development of such inhomogeneities highlights a fundamental departure from the assumptions underlying classical kinetic theory for molecular gases, where collisions conserve energy and homogeneous states are typically stable. In granular systems, even weak inelasticity is sufficient to destabilize a uniform state and drive the spontaneous emergence of spatial structure. The resulting correlations and heterogeneous particle distributions play a crucial role in determining the macroscopic dynamics of granular systems and have important implications for aggregation processes, since collision rates and particle encounters may become strongly influenced by local clustering rather than by homogeneous mixing.

The emergence of clustering and spatial correlations in granular systems has important consequences for the kinetic description of aggregation processes. Classical coagulation theory, as embodied in the Smoluchowski equation, assumes that particles are well mixed and that collision rates can be expressed in terms of mean-field kernels depending only on particle properties and relative velocities. However, in dissipative particulate systems the development of density fluctuations and correlated particle motion may substantially modify the effective collision rates~\cite{BrilliantovPoschel2004}. Regions of enhanced particle concentration can experience elevated collision frequencies, while dilute regions may contribute little to the overall aggregation dynamics. As a result, the assumption of homogeneous mixing underlying many coagulation models may break down when clustering becomes significant~\cite{leyvraz2003scaling}. From this perspective, the collision kernel appearing in kinetic aggregation equations should be understood not only as a function of microscopic interaction physics, but also as reflecting the evolving spatial structure of the particle ensemble. In the following sections we explore how these granular effects can be incorporated into coagulation frameworks and how modern computational approaches provide a powerful route for investigating this interplay between aggregation kinetics and granular dynamics.

\subsection{Bridging coagulation kinetics and granular physics}

The correspondence between coagulation theory and granular kinetic theory
can be made explicit by comparing the key quantities that govern particle
interactions in each framework. In the Smoluchowski description, aggregation
kinetics are controlled by the collision kernel $\beta_{ij}$ and the
sticking probability $\alpha_{ij}$, which together determine the rate at
which particles of sizes $i$ and $j$ form larger aggregates. In granular
kinetic theory, the analogous quantities are the collision frequency
between particles and the coefficient of restitution, which characterizes
the degree of energy dissipation during impact. Finally, the mean-field assumption underlying the Smoluchowski equation corresponds most closely to the dilute limit of granular kinetic theory, where particle correlations are weak and collisions can be treated as binary and uncorrelated events. These analogies highlight how the two frameworks can be viewed as complementary descriptions of interacting particulate systems, differing primarily in the emphasis placed on aggregation outcomes versus collisional dynamics. The main
correspondences between the two descriptions are summarized in
Table~\ref{tab:framework_mapping}.

\begin{table}[t]
\centering
\caption{Conceptual correspondence between the Smoluchowski coagulation framework and granular kinetic theory.}
\label{tab:framework_mapping}
\begin{tabular}{lL{4.0cm}L{4.0cm}}
\toprule
\textbf{Coagulation theory} & \textbf{Granular physics} & \textbf{Physical interpretation} \\
\midrule

Collision kernel $\beta_{ij}$ & Collision frequency & Rate of particle encounters \\

Sticking probability $\alpha_{ij}$ & Coefficient of restitution $e$ & Dissipation controls collision outcome \\

Mean-field approximation & Dilute granular gas limit & Binary, uncorrelated collisions \\

Particle concentration $n_i$ & Conjoint Position and Velocity distribution $f(\mathbf{r},\mathbf{v})$ & Statistical description of particles \\

Aggregation rate & Collisional energy dissipation & Microscopic dynamics driving evolution \\

\bottomrule
\end{tabular}
\end{table}

While these correspondences provide a useful conceptual bridge between
coagulation and granular kinetic theories, the analogy becomes
increasingly limited when the assumptions underlying the mean-field
description are no longer satisfied. In particular, at sufficiently high
collision rates the probability that particles undergo multiple
interactions over short time intervals increases, leading to correlations
between particle velocities and positions that violate the assumption of
uncorrelated binary collisions~\cite{vanNoijeErnst1998, Brey1998}. In this regime, the effective collision
rates may deviate significantly from those predicted by mean-field kernels.
A second limitation arises at high densities, where the
dissipative nature of collisions promotes the formation of clusters.
Because particles in dense regions experience more frequent collisions and
lose kinetic energy more rapidly, they tend to remain locally confined,
giving rise to clustering instabilities and long-lived density
inhomogeneities~\cite{GoldhirschZanetti1993,Goldhirsch2003}. Energy dissipation also affects the velocity statistics of
the system, producing non-Maxwellian velocity distributions and spatial
variations in the granular temperature~\cite{BrilliantovPoschel2004, PhysRevLett.95.068001}. These effects further modify local
collision rates and aggregation probabilities. As a result, granular
systems often develop pronounced spatial inhomogeneities in which the
aggregation dynamics become controlled by localized structures rather than
by homogeneous mixing~\cite{Goldhirsch2003,PhysRevLett.74.4114}. In such situations, the effective collision kernels
cannot be described solely by particle properties and relative velocities,
but must also account for the evolving spatial correlations within the
particle ensemble.

The complexity of aggregation dynamics, together with the limitations of
analytical solutions, has led to the widespread use of computational
approaches for studying coagulation processes in particulate systems.
At the continuum level, the evolution of the particle size distribution
can be obtained by solving the Smoluchowski equation numerically, either
through direct integration of the rate equations or through sectional
and moment-based methods that approximate the evolving distribution
\cite{friedlander2000smoke,leyvraz2003scaling}. These approaches are computationally
efficient and particularly well suited for regimes in which the
mean-field assumptions of homogeneous mixing and uncorrelated binary
collisions remain valid. However, when spatial correlations,
clustering, or complex collision physics become important, particle-resolved methods provide a more faithful description of the system.
Discrete element modeling (DEM) and molecular dynamics (MD) simulations
explicitly track the trajectories and interactions of individual grains,
allowing the incorporation of detailed contact mechanics, energy
dissipation, and spatial structure \cite{CundallStrack1979,BrilliantovPoschel2004}.
Between these two limits, hybrid approaches combine continuum
descriptions of particle populations with particle-based treatments of
local interactions, thereby retaining key microscopic physics while
maintaining computational efficiency~\cite{sen2014multi, patterson2012stochastic}. The choice between continuum,
discrete, or hybrid methods therefore depends largely on the physical
regime of interest: continuum models are most appropriate for dilute,
well-mixed systems, whereas particle-based simulations become essential
when aggregation dynamics is strongly influenced by correlations and
spatial inhomogeneities.

\section{Collision Dynamics: Sticking, Bouncing, and Fragmentation}
\begin{figure}[h!]
    \centering
    \includegraphics[width=1.0\linewidth]{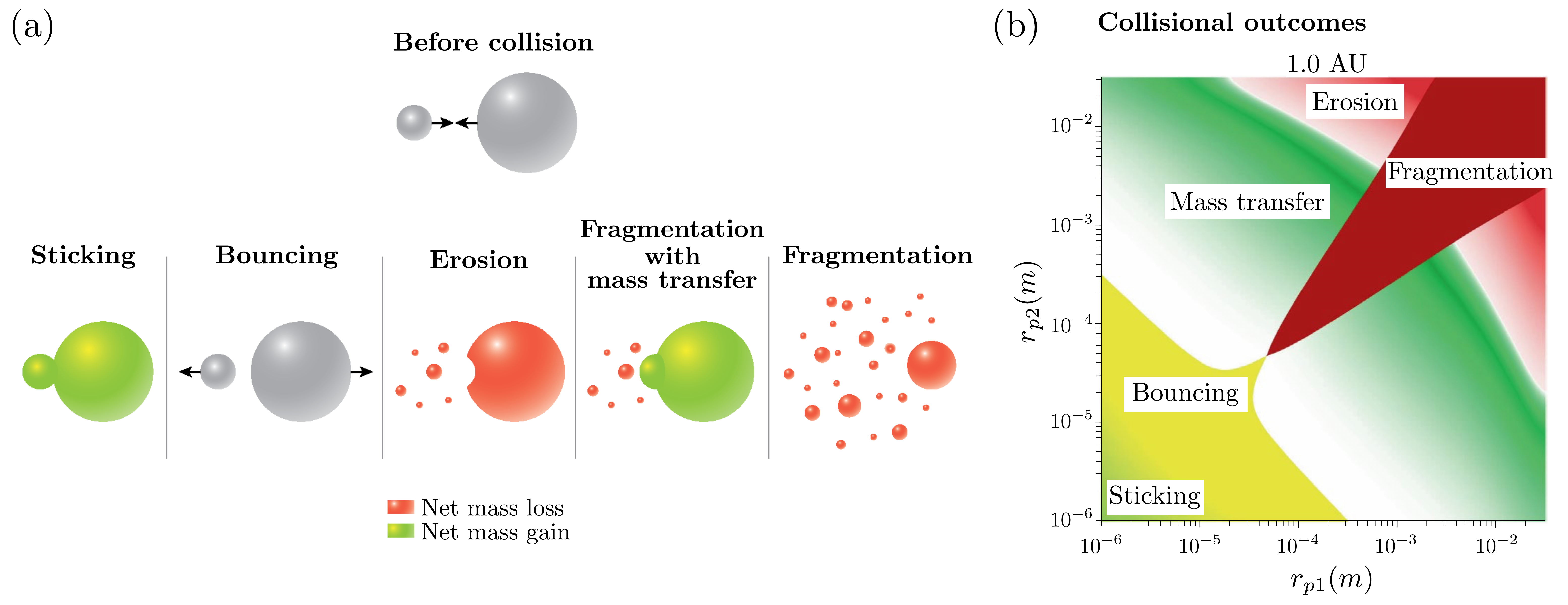}
    \caption{
Collision outcomes for particle aggregates. 
(a) Schematic representation of the main regimes of collisional
interactions between dust aggregates, including sticking, bouncing,
erosion, and fragmentation, as a function of collision velocity and
aggregate properties. Adapted from \cite{Birnstiel2024}. 
(b) Example phase diagram illustrating the regions of parameter space in
which different collision outcomes occur, shown here as a function of the
size ratio between colliding particles. This diagram is derived from
modelling of dust evolution in the protoplanetary disk around the star
LkCa 15 \cite{Husmann2016}. Laboratory and microgravity experiments have
also mapped these regimes under controlled conditions, providing
empirical constraints on the transitions between sticking, bouncing,
erosion, and fragmentation \cite{Dominik1997,Blum2008,Guttler2010,Blum2018}.
}
    \label{fig:collision_outcomes}
\end{figure}

Collisions between particles do not always lead to aggregation, and the
outcome of an encounter depends sensitively on the physical conditions of
the collision. In particular, parameters such as the impact velocity,
particle mass and size ratio, internal structure of the aggregates, and
material properties of the grains determine whether particles stick,
bounce, erode, or fragment upon collision. A schematic overview of these
different collision outcomes is shown in
Fig.~\ref{fig:collision_outcomes}(a), adapted from \cite{Birnstiel2024}.
In addition to this qualitative classification, a number of studies have
attempted to map the regions of parameter space where each regime
dominates. Figure~\ref{fig:collision_outcomes}(b) presents an example of
such a phase diagram derived from modeling the dust population in the
protoplanetary disk around the star LkCa 15 \cite{Husmann2016}. Similar
collision maps have also been obtained from laboratory and microgravity
experiments that investigate dust aggregate collisions under controlled
conditions. These experimental studies have systematically explored the
transition between sticking, bouncing, erosion, and fragmentation as a
function of impact velocity, particle size, and aggregate structure
\cite{Dominik1997,Blum2008,Guttler2010,Lee2015,Blum2018}. Together, such
experimental and modeling efforts provide important constraints on the
collision physics that governs the growth and evolution of particulate
systems.

\subsection{Collision Outcome Regimes}
In the low-velocity regime, collisions between particles may lead to
\textit{sticking}, allowing aggregates to grow through successive
coagulation events. Several physical mechanisms can provide the
attractive forces required to stabilize contacts between colliding
grains. Van der Waals forces are always present and dominate adhesion
between small particles, particularly for micron-sized dust grains where
surface forces exceed the kinetic energy of impacts
\cite{Dominik1997}. This
effect is especially pronounced for ice-mantled grains, since water ice
has a specific surface energy roughly an order of magnitude higher than
that of silicates ($\sim 0.19$~J\,m$^{-2}$~\cite{Gundlach2011}),
substantially raising the sticking threshold of icy aggregates even at
temperatures where no liquid or quasi-liquid phase is present~\cite{Gartner_2017}. In environments where particles carry electric
charges, electrostatic interactions may also contribute to attractive
forces between grains \cite{Lee2015}. Additional mechanisms may operate under specific
environmental conditions; at higher temperatures,
sintering processes can create solid bonds between grains through
thermally activated surface diffusion, stabilizing aggregates over
longer timescales. For all these mechanisms, sticking generally occurs
only below a critical impact velocity, above which the kinetic energy of
the collision exceeds the energy that can be dissipated through contact
deformation or surface bonding. Experimental studies have played a key
role in identifying these velocity thresholds and characterizing the
conditions under which sticking transitions to bouncing or
fragmentation \cite{Blum2008,Guttler2010,Lee2015,Blum2018,Poppe_2000,Gundlach_2015}.

At intermediate collision velocities, particle encounters often lead to
\textit{bouncing} rather than sticking. In this regime the kinetic energy of the collision is sufficiently large to prevent the formation of stable
adhesive contacts, yet not high enough to cause fragmentation of the
aggregates. Instead, particles rebound after impact, possibly undergoing
minor restructuring or compaction during the brief contact phase. The
transition from sticking to bouncing typically occurs once the kinetic
energy of the collision exceeds the energy that can be dissipated through
surface forces and internal deformation. As aggregates grow in size, their
relative velocities generally increase due to processes such as turbulent
motion, differential settling, or radial drift, making bouncing collisions
increasingly likely. This behaviour gives rise to the so-called
\textit{bouncing barrier}, a regime in which aggregates repeatedly collide
and rebound without significant net growth~\cite{Guttler2010,zsom2010outcome}. Laboratory experiments and
numerical studies have shown that porous dust aggregates can become
compacted through repeated bouncing collisions, which further reduces the
efficiency of energy dissipation and suppresses sticking in subsequent
impacts~\cite{blum2000,Dominik1997,Okuzumi2009,Tatsuuma_2023,Weidling_2009}. The bouncing barrier therefore represents a major obstacle for
continued particle growth in many aggregation models, particularly in
protoplanetary disks where particles may reach millimetre or centimetre
sizes before further growth becomes inefficient~\cite{wurm2021understanding-87f}. Several mechanisms have been proposed to overcome the bouncing barrier
and enable continued particle growth. One promising possibility is the
contact electrification of dust grains, which can generate long-range
electrostatic forces that increase the effective attraction between
particles and promote sticking at collision velocities that would
otherwise result in bouncing \cite{Lee2015,steinpilz2020electrical}.

\begin{figure}[h!]
    \centering
    \includegraphics[width=1.0\linewidth]{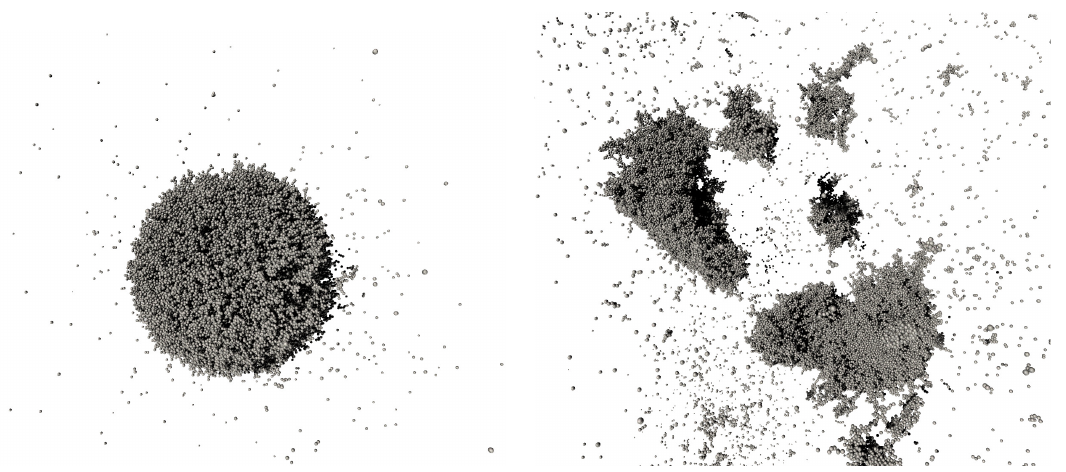}
    \caption{Transition in collision outcomes as a function of projectile size. As the mass ratio increases, the system shifts from a regime of surface erosion to fragmentation.~\cite{Seizinger2013}}
    \label{fig:fractals}
\end{figure}
At sufficiently high collision velocities, particle impacts may lead to
\textit{erosion} or \textit{fragmentation}, marking the transition to a regime
in which aggregates lose mass rather than grow~\cite{Wurm2005, Seizinger2013, teiser2009highvelocity-074}. In fragmentation
collisions, the kinetic energy of the impact exceeds the binding energy
of the aggregate, causing it to break into a distribution of smaller
fragments. The threshold velocity for fragmentation depends strongly on
the material properties of the grains, the porosity of the aggregates,
and the size ratio of the colliding particles. For highly porous dust
aggregates composed of micron-sized monomers, laboratory experiments
suggest fragmentation thresholds on the order of a few meters per second,
although significant variations occur depending on composition and
structure \cite{Blum2008,Guttler2010}. In addition to catastrophic
fragmentation, collisions involving a small projectile and a much larger
target can lead to \textit{erosion}, where material is gradually removed
from the surface of the larger aggregate through repeated impacts~\cite{Seizinger2013,becker2022releasing-520,becker2023ejected-cea,Schrapler_2011,Gundlach_2015} (Fig.~\ref{fig:fractals}). Under
certain conditions, however, fragmentation events may also involve
\textit{mass transfer}, in which part of the projectile material remains
attached to the target despite partial breakup. Such processes can allow
continued growth even in regimes where fragmentation becomes important,
and they have been proposed as possible pathways for overcoming growth
barriers in protoplanetary disks \cite{Blum2018}.

Collisions between aggregates may also lead to significant internal
\textit{compaction and restructuring}, even when sticking or bouncing
remains the dominant outcome. During an impact, the kinetic energy of the
collision can be partially dissipated through the rearrangement of
constituent grains within the aggregate, causing a reduction in porosity
and an increase in packing density. Repeated collisions therefore tend
to progressively compact initially fluffy aggregates, altering their
internal structure and mechanical properties
\cite{Dominik1997,Blum2008}. As a consequence, the morphology of
growing aggregates evolves over time, often characterized by an
increasing fractal dimension that reflects the gradual transition from
highly porous clusters to more compact structures. This structural
evolution plays an important role in determining the aerodynamic and
mechanical properties of aggregates~\cite{paszun2009collisional-d05}. However, such effects are difficult
to capture within classical Smoluchowski-type coagulation models, where
the fractal dimension is often introduced as a fixed parameter that
prescribes the relation between aggregate mass and size rather than
emerging dynamically from the growth process, although dynamic formulations have been proposed~\cite{1993AA280617O}. In contrast, particle-based
simulations can explicitly resolve the restructuring of aggregates and
track the time-dependent evolution of their morphology~\cite{wada2007numerical-920,wada2008numerical-9a9, KEMPF1999388}. The compaction
history of an aggregate may also introduce a form of \textit{mechanical
memory}, whereby previous collisions influence the response to future
impacts~\cite{Dominik1997, paszun2009collisional-d05, PhysRevLett.93.115503}. These effects ultimately determine the strength, density, and
stability of aggregates, with important implications for their subsequent
collisional evolution.

\subsection{Collision Kernels from a Granular Perspective}
In the Brownian regime, particle collisions are primarily driven by
thermal motion arising from interactions with the surrounding fluid. The
random fluctuations in particle velocities lead to diffusive transport
and ultimately to collisions between particles of different sizes. In
this limit, the classical Smoluchowski theory predicts a collision kernel
that depends on the sum of particle radii and the thermal diffusivities
of the colliding species, leading to a rate that scales approximately as
$K_{ij} \propto (r_i + r_j)$ for small particles suspended in a gas or
liquid \cite{smoluchowski1916drei,friedlander2000smoke}. Because the collision
velocities are typically small in this regime, Brownian coagulation is
particularly effective for submicron particles whose motion is strongly
coupled to the thermal fluctuations of the medium. From a granular
perspective, however, the outcome of such collisions may still depend on
microscopic dissipation mechanisms and surface forces. In particular, the
probability that two particles remain attached after contact can depend
on the effective temperature of the system, which controls the magnitude
of velocity fluctuations and the ability of particles to overcome
adhesive interactions. These considerations introduce a temperature-dependent sticking probability that modifies the effective coagulation
kernel~\cite{Dominik1997}. The Brownian regime therefore represents a limit in which
thermal agitation dominates particle dynamics, and it is most relevant
for very small particles, typically with sizes in in the nanometer to micrometer size range.

At larger particle sizes or in environments where thermal agitation
becomes negligible, collisions are primarily driven by inertial motion
rather than Brownian diffusion. In this \textit{ballistic} regime,
particles follow approximately straight trajectories between collisions,
and the collision rate is determined by the geometric cross section and
the relative velocity of the interacting particles. The relevant
velocity scale is no longer set by thermal fluctuations but instead by
macroscopic dynamical processes such as gravitational settling,
differential drift, turbulent fluctuations, or externally driven flows.
As particle sizes increase, these mechanisms can produce relative
velocities significantly larger than those arising from Brownian motion,
leading to collisions with higher impact energies. In granular systems,
the outcome of such collisions depends sensitively on dissipative
contact mechanics, aggregate structure, and the velocity-dependent
sticking probability. Consequently, the effective collision kernel may
deviate from the purely geometric prediction when bouncing,
fragmentation, or compaction become important. This regime is
particularly relevant for millimeter- to centimeter-sized aggregates in
protoplanetary disks, as well as for dense particulate systems where
particle inertia dominates over thermal motion
\cite{Safronov1972,Blum2008}.

In many natural and industrial environments, particle collisions are
strongly influenced by turbulent fluctuations in the surrounding fluid.
In this regime, turbulent eddies generate relative velocities between
particles through velocity gradients and inertial effects, leading to
enhanced collision rates compared to purely Brownian motion. For small
particles that closely follow the fluid motion, collisions can arise
from local shear within turbulent eddies, an effect first described by
Saffman and Turner \cite{SaffmanTurner1956}. As particle sizes increase,
inertial effects become more important, and particles may decouple
partially from the flow, leading to additional sources of relative
velocity such as preferential concentration and crossing trajectories.
These mechanisms can significantly increase the frequency of particle
encounters and promote clustering within turbulent flows
\cite{Falkovich2002}. From a granular perspective, turbulence therefore
not only enhances collision rates but also modifies the spatial
distribution of particles, generating localized regions of high
concentration where aggregation may proceed more rapidly. Such effects
introduce correlations between particle velocities and positions that
are not captured by mean-field coagulation models. Turbulence-driven
collisions are particularly important for intermediate particle sizes,
where Brownian motion becomes weak while particle inertia remains
insufficient to dominate the dynamics entirely.

Taken together, these regimes illustrate how the classical collision
kernels of coagulation theory can be reinterpreted through the lens of
granular dynamics. While Brownian, inertial, and turbulence-driven
collisions provide the physical mechanisms that set the encounter rate
between particles, the effective aggregation rate is ultimately
controlled by additional processes characteristic of granular systems. The effective collision
kernel may depend not only on particle size and relative velocity, but
also on the evolving structure and spatial correlations within the
particle population. Incorporating these effects remains an important
challenge for kinetic descriptions of aggregation and highlights the
need to combine classical coagulation theory with insights from
granular physics.

\subsection{Scaling Behaviour and Dynamical Classes of Coagulation Kernels}

An important insight into the behaviour of aggregation processes can be
obtained by examining the scaling properties of the collision kernel.
Many physically relevant kernels can be approximated as homogeneous
functions of the cluster masses, such that
\begin{equation}
K(\xi\,i,\xi\,j)=\xi^{\lambda}K(i,j), \qquad \xi>0,
\end{equation}where $\lambda$ denotes the homogeneity exponent
\cite{leyvraz2003scaling}. This parameter plays a central role in determining
the large-scale dynamics of the aggregation process. In particular,
kernels with sufficiently large values of $\lambda$ can lead to
\textit{gelation}, a runaway growth phenomenon in which a finite fraction
of the system mass condenses into a macroscopic cluster within finite
time $t_{\mathrm{c}}$. The onset of gelation is signalled by the
divergence of the second moment of the distribution,
\begin{equation}
M_2(t)=\sum_k k^2\, n_k(t)\;\longrightarrow\;\infty
\qquad \text{as } t\to t_{\mathrm{c}}^{-},
\end{equation}
which measures the weight-averaged cluster mass and remains finite for
all $t<t_{\mathrm{c}}$. At the gel point the distribution itself becomes
scale-free, decaying as a pure power law
\begin{equation}
n_k(t_{\mathrm{c}})\sim k^{-\tau}, \qquad \tau=\frac{3+\lambda}{2}.
\end{equation}
A defining feature of the gelling regime is that mass is no longer
conserved within the population of finite clusters once $t>t_{\mathrm{c}}$:
the ``missing'' mass is continuously transferred from the sol to the gel
\cite{leyvraz2003scaling,dongen1988scaling-b3e}. The
product kernel $K(i,j)=ij$ ($\lambda=2$) is the canonical example of
gelation and reproduces the mean-field exponent $\tau=5/2$ familiar from
Flory--Stockmayer polymerisation theory~\cite{Stockmayer1944}. In the astrophysical context,
gelation provides a natural description of \textit{runaway growth}, in
which a small number of clusters rapidly decouple from the background
population and accumulate a dominant share of the available mass.

\begin{figure}[h!]
    \centering
    \includegraphics[width=0.45\linewidth]{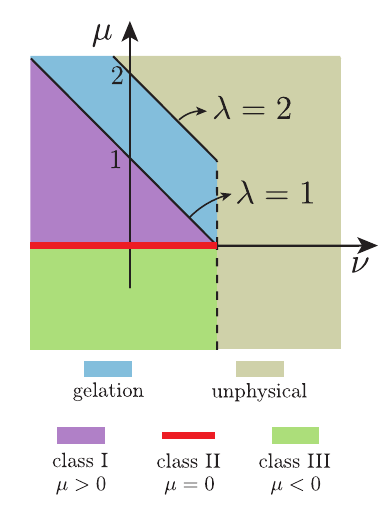}
    \caption{Schematic phase diagram of homogeneous coagulation kernels in
the $(\nu,\mu)$ plane. For strongly asymmetric collisions ($i\ll j$) the
kernel scales as $K(i,j)\sim i^{\mu}j^{\nu}$, and the homogeneity
exponent is $\lambda=\mu+\nu$; the solid diagonal lines are loci of
constant $\lambda$. All three exponents are dimensionless. The sign of
$\mu$ defines the three dynamical classes of mass-conserving aggregation
($\lambda\le 1$): class~I ($\mu>0$), class~II ($\mu=0$), and class~III
($\mu<0$). The solid line $\lambda=1$ marks the boundary above which
runaway growth (gelation) occurs, while $\lambda=2$ (the product kernel
$K\propto ij$) sets the upper limit of physically admissible homogeneity;
the dashed line $\nu=1$ separates this region from the unphysical regime
($\nu>1$ or $\lambda>2$). Adapted from \cite{ernst1986fractals-a2a}.}
    \label{fig:kernel_classes}
\end{figure}

While the exponent $\lambda$ controls the overall scaling behaviour of
the kernel, additional properties of the cluster size distribution
$n_k(t)$ depend on how the kernel behaves when collisions involve
particles of very different sizes. This behaviour can be characterized
by a second exponent $\mu$, defined through the asymptotic relation
\begin{equation}\label{eq:K_homog}
K(i,j) \sim i^{\mu} j^{\lambda-\mu}, \qquad i \ll j .
\end{equation}
Depending on the value of $\mu$, three broad classes of aggregation
kernels can be distinguished.

Class~I corresponds to $\mu>0$, where collisions between clusters of
comparable size dominate the growth process. This behaviour is typical
of reaction-limited aggregation processes, such as polymerization
reactions. Class~III corresponds to $\mu<0$, where growth is primarily
driven by collisions between large clusters and much smaller particles.
Diffusion-limited aggregation processes, such as Brownian coagulation
in aerosols, provide representative examples of this class. The
intermediate case $\mu=0$ defines class~II, which forms the boundary
between the two regimes and corresponds to situations in which no
single growth mechanism clearly dominates the dynamics.

These different dynamical classes are summarized schematically in
Fig.~\ref{fig:kernel_classes}, which illustrates the phase diagram of
aggregation kernels in the $(\lambda,\mu)$ parameter space. The diagram
highlights the qualitative changes in aggregation behaviour that arise
from different scaling properties of the collision kernel, including
the transition to runaway growth in the gelation regime.


\section{Growth Kinetics and Aggregation Regimes}\label{sec:sec4}

Having established how the physical mechanisms of particle collisions
determine the form of the coagulation kernel, we now turn to the
implications of these kernels for the temporal evolution of particle
populations. In the Smoluchowski framework, the collision kernel
encodes the microscopic physics of particle encounters, while the
resulting kinetic equation governs the evolution of the cluster size
distribution $n_k(t)$. Different kernels can therefore produce
qualitatively different growth dynamics, ranging from smooth
self-similar evolution to runaway growth and gelation. In many
aggregation systems, the particle size distribution evolves toward a
scaling regime in which the distribution retains a universal shape
while a characteristic mass scale grows in time. The rate at which this
characteristic mass increases depends on the scaling properties of the
collision kernel, as discussed in the previous section. However, real
particulate systems often exhibit additional complexities that are not
fully captured by idealized mean-field models. Dissipative collisions,
aggregate restructuring, and spatial correlations can all influence the
effective aggregation rates and modify the growth kinetics. As a
result, the evolution of particle populations may deviate from the
simplest scaling predictions, particularly in systems where clustering,
fragmentation, or growth barriers become important.

\subsection{Temporal Evolution and Self-Similar Growth}

The temporal evolution of particle populations in aggregation systems is
often characterized by the emergence of self-similar behavior. In this
regime, the cluster size distribution evolves in such a way that its
overall shape remains invariant when rescaled by a characteristic mass
or size scale that grows with time. Formally, this behavior can be
expressed through a scaling ansatz of the form
\begin{equation}
n_k(t) \sim s(t)^{-2} \, \Phi\!\left(\frac{k}{s(t)}\right),
\end{equation}
where $k$ is the cluster size (number of monomers per cluster), \(s(t)\) denotes the characteristic cluster size and
\(\Phi\) is a universal scaling function that depends only on the ratio
\(k/s(t)\) \cite{leyvraz2003scaling}. The growth rate of \(s(t)\) is controlled
primarily by the homogeneity exponent \(\lambda\) of the collision
kernel. For kernels with \(\lambda < 1\), the characteristic mass grows
smoothly in time and the system approaches a self-similar scaling
state. In contrast, kernels with sufficiently large homogeneity
exponents can lead to runaway growth or gelation, in which a finite
fraction of the total mass accumulates in a macroscopic cluster within
finite time. Although these scaling solutions provide a powerful
framework for understanding aggregation kinetics, real particulate
systems often deviate from the idealized mean-field predictions.
Processes such as fragmentation, compaction, and spatial clustering can
modify the effective collision rates and alter the temporal evolution
of the particle population.

As the system evolves, the self-similar regime does not necessarily persist indefinitely. Deviations from scaling can arise when the coagulation kernel increasingly favours interactions involving the largest clusters, leading to a progressive acceleration of mass transfer toward the high-mass tail of the distribution. In such situations, growth becomes dominated by a small subset of rapidly evolving aggregates that act as efficient collectors of surrounding particles. This transition reflects a fundamental change in the hierarchy of collision events: instead of interactions occurring predominantly between clusters of comparable size, collisions involving highly disparate masses begin to dominate the dynamics.\\
\indent From a kinetic perspective, this behaviour emerges naturally in kernels that grow sufficiently rapidly with cluster size. Such kernels enhance the probability that large aggregates sweep up smaller ones, amplifying fluctuations in cluster mass and driving the system away from the self-similar scaling regime. Similar mechanisms appear in a variety of physical systems, including aerosol aggregation, gravitational clustering, and particulate growth in turbulent environments. In these cases, the redistribution of mass toward a small number of dominant clusters provides a useful conceptual framework for understanding how microscopic interaction rules translate into large-scale structural evolution~\cite{lushnikov2006gelation-506}. 

Such behaviour is of particular
interest in astrophysical contexts, where runaway growth of large
planetesimals can occur once gravitational focusing enhances the
collision cross section of the largest bodies~\cite{KOKUBO200015}. The collision cross section is enhanced over its
geometric value by the gravitational focusing factor,
\begin{equation}
\sigma = \pi R^2\left(1+\frac{v_{\rm esc}^2}{v^2}\right),
\end{equation}
where $v_{\rm esc}=\sqrt{2Gm/R}$ is the surface escape velocity and $v$
the relative velocity of the colliding bodies. Since
$v_{\rm esc}^2\propto m/R \propto m^{2/3}$, the focusing term grows with
mass, so that the largest bodies accrete disproportionately faster than
the rest of the population. In the language of Sect.~3c, this renders
the effective collision kernel superlinear in mass, with a homogeneity
exponent $\lambda>1$ (see Eq.~\ref{eq:K_homog}). Runaway planetesimal growth is
therefore the astrophysical counterpart of the gelation transition: a
small number of bodies rapidly decouple from the size distribution and
accumulate a dominant fraction of the total mass, in direct analogy with
the formation of a gel~\cite{KOKUBO200015}.  Similar phenomena also
arise in polymerization and other chemical aggregation processes,
where the formation of a giant cluster dramatically alters the
subsequent dynamics of the system \cite{leyvraz2003scaling,ernst1986fractals-a2a}.
In realistic particulate systems, the onset and nature of runaway
growth may be further modified by additional effects such as
fragmentation, compaction, or spatial clustering, which can either
delay or suppress the formation of a macroscopic aggregate.

\subsection{Growth Barriers and Collision Regimes}

While runaway growth and gelation provide a natural outcome of coagulation dynamics in many theoretical models, real particulate systems often encounter mechanisms that hinder or delay aggregation. Laboratory experiments and observations in natural environments consistently show that particle growth does not always proceed as rapidly as predicted by idealized kernels~\cite{friedlander2000smoke, dominikh2007,blum2000,Emets1994}. Instead, a variety of physical effects can reduce the probability that collisions lead to successful sticking. Electrostatic charging, surface forces, particle restructuring, and collisional rebound may all act to suppress effective coagulation even when encounter rates remain high~\cite{matthews2004effects-8d8}. Similar obstacles are well documented in astrophysical environments, where the growth of dust grains in protoplanetary disks is limited by mechanisms such as repulsive charging, bouncing collisions, and fragmentation at sufficiently high impact velocities~\cite{Blum2008,dominik2006growth,guttler2010outcome-dfe}. These effects highlight an important limitation of purely kinetic descriptions of aggregation: in many real systems, the rate of cluster growth reflects not only how frequently particles collide, but also whether those collisions result in stable aggregates. Understanding these barriers is therefore essential for connecting theoretical predictions of runaway growth with experimentally and observationally accessible regimes.

One of the most widely studied mechanisms capable of suppressing aggregation is electrostatic charging. When insulating particles collide or interact with surrounding media, they may acquire electric charge through processes such as triboelectric charging, plasma interactions, or photoelectric emission. The resulting electrostatic Coulomb forces can significantly modify collision dynamics, particularly for small particles where electrostatic interactions may exceed thermal or gravitational forces~\cite{friedlander2000smoke,Lee2015,matthews2008charging, okuzumi2009electric}. If particles acquire charges of the same sign, the resulting repulsive potential creates an energetic barrier that reduces the probability of successful collisions. As a concrete example of such a modification, the Brownian collision
kernel can be multiplied by a dimensionless factor that accounts for the
electrostatic interaction between charged particles,
\begin{equation}
\beta_{ij} = \beta_{ij}^{\mathrm{B}}\, W_{ij},
\qquad
W_{ij} = \frac{\kappa_{ij}}{e^{\kappa_{ij}} - 1},
\qquad
\kappa_{ij} = \frac{q_i q_j}{4\pi\varepsilon_0 (r_i + r_j)\, k_{\mathrm{B}} T},
\end{equation}
where $\beta_{ij}^{\mathrm{B}}$ is the neutral Brownian kernel, $q_i$ and
$q_j$ are the particle charges, and $\kappa_{ij}$ is the ratio of the
Coulomb interaction energy at contact to the thermal energy. The factor
$W_{ij}$ reduces the collision rate for like-charged particles
($\kappa_{ij} > 0$) and enhances it for oppositely charged ones
($\kappa_{ij} < 0$), recovering the neutral kernel as
$\kappa_{ij} \to 0$. This form, introduced for charged microparticles in
a plasma by Ivlev et al.~\cite{ivlev2002coagulation-51a}, illustrates how long-range
interactions are incorporated into the kernel without altering the
structure of the Smoluchowski equation itself. In such situations, coagulation rates may become orders of magnitude smaller than those predicted by neutral-particle kernels. Electrostatic effects are known to play an important role in a variety of systems, including atmospheric aerosols, dusty plasmas, and protoplanetary dust environments. In these contexts, particle charging not only modifies collision cross sections but may also introduce long-range interactions that alter the structure and morphology of the resulting aggregates. Incorporating electrostatic interactions into coagulation models therefore represents an important step toward bridging the gap between idealized kinetic descriptions and experimentally observed aggregation dynamics \cite{ourPRE}.

Interestingly, electrostatic interactions do not always suppress aggregation. Even when particles carry net charges of the same sign, polarization effects may introduce attractive contributions to the interaction potential. When a charged particle approaches a neutral or weakly charged aggregate, it can induce a dipole moment in the neighbouring particle, generating a charge--dipole interaction that enhances the effective collision cross section. Similar effects arise between charged aggregates when their internal charge distributions become polarized by the electric field of nearby particles. These induced interactions can partially compensate for Coulomb repulsion and, in some cases, even lead to net attractive forces at short distances. As a result, the effective coagulation kernel may differ substantially from that predicted by purely geometric or purely Coulombic considerations. Such polarization-driven interactions have been observed or inferred in a range of systems including dusty plasmas, aerosol particles, and granular materials. In the astrophysical context, microgravity and laboratory
experiments have shown that collisional (tribo-)charging can drive the
growth of charged grains into much larger aggregates than sticking forces
alone would allow, bridging the classical bouncing barrier in
protoplanetary disks \citep{steinpilz2020electrical,Teiser2025}; in
particular, Teiser et al. follow the growth of tribocharged
grains up to a centimetre-scale impact-erosion limit. Incorporating these effects into aggregation models therefore provides a more realistic description of particle growth in environments where electrostatic charging and dielectric response play an important role~\cite{Teiser2025}.

Recent studies also suggest that the statistical distribution of particle charge can play an important role in determining aggregation kinetics~\cite{ourPRE}. In particular, charge distributions with pronounced heavy tails may significantly enhance coagulation rates compared to the more commonly assumed Gaussian statistics. In such systems, a small fraction of particles can acquire unusually large charges, producing strong electrostatic interactions that modify collision probabilities and accelerate cluster growth. Numerical simulations indicate that this effect may generate an intermediate-time regime of rapid aggregation in which the presence of highly charged particles effectively promotes the formation of larger clusters. The resulting aggregates are expected to possess stronger electrostatic cohesive energies, potentially making them more resistant to external perturbations such as shear, turbulence, or collisional fragmentation. From a kinetic perspective, heavy-tailed charge statistics therefore provide a natural pathway for the early formation of large aggregates before other mechanisms capable of limiting growth become dominant. These results highlight how fluctuations in microscopic particle properties can strongly influence the macroscopic evolution of coagulating systems.

In classical coagulation theory, collisions are typically assumed to lead to irreversible sticking, so that every encounter between particles results in the formation of a larger aggregate. Experimental studies of particulate systems, however, reveal that collision outcomes can be considerably more complex~\cite{blum2000,guttler2010outcome-dfe}. Depending on the relative velocity, impact parameter, and internal structure of the aggregates, collisions may result not only in sticking but also in bouncing, partial restructuring, or fragmentation. In many cases, energy dissipation during contact plays a central role in determining the outcome of an encounter. When sufficient kinetic energy is dissipated through deformation, rolling, or sliding at the contact points between constituent particles, aggregates may remain bound and continue to grow. Conversely, if dissipation is insufficient, colliding clusters may rebound without forming a stable aggregate, effectively reducing the sticking probability. At higher impact energies, fragmentation processes may dominate, producing smaller fragments and altering the cluster size distribution~\cite{Brilliantov2001, akimkin2023coagulationfragmentation-f58}. These competing outcomes can be incorporated into the kinetic description
by augmenting the Smoluchowski equation with a fragmentation term that
counteracts aggregation. Adding such a binary break-up contribution to
Eq.~(\ref{eq:smoluchowski}) yields the coagulation--fragmentation equation,
\begin{equation}
\frac{dn_i}{dt}
= \underbrace{\frac{1}{2}\sum_{j=1}^{i-1}\beta_{j,\,i-j}\,n_j n_{i-j}
  - n_i\sum_{j=1}^{\infty}\beta_{ij}\,n_j}_{\text{aggregation}}
\;\;\underbrace{-\;\frac{1}{2}\,a_i n_i
  + \sum_{j=1}^{\infty} a_{i+j}\,b_{i,\,i+j}\,n_{i+j}}_{\text{fragmentation}},
\end{equation}
where $a_i$ is the rate at which clusters of size $i$ break up and
$b_{i,j}$ is the average number of fragments of size $i$ produced by the
break-up of a cluster of size $j$. Unlike pure coagulation, this
equation does not necessarily drive the system toward unbounded growth:
the competition between aggregation and break-up can instead select a
stationary cluster-size distribution, providing a natural description of
systems that reach a dynamic balance between growth and disruption (see Sec.~\ref{sec:sec4}\ref{sec:saturn}). These mechanisms introduce an additional layer of complexity into coagulation dynamics, as the effective aggregation rate becomes controlled not only by collision frequencies but also by the detailed mechanics of particle contacts.

The physical mechanisms discussed above illustrate that the aggregation dynamics of real particulate systems are shaped by a combination of collision frequencies, interaction forces, and contact mechanics. While classical Smoluchowski theory provides a powerful framework for describing coagulation through a kernel that depends on particle properties such as size or mobility, many experimentally relevant processes introduce additional layers of complexity. Electrostatic charging, polarization effects, statistical variability in particle charge, and the diversity of collision outcomes all modify either the probability that particles encounter one another or the likelihood that collisions lead to stable aggregates. As a result, the effective aggregation kinetics often depart from those predicted by idealized kernels. A common strategy to account for these effects is to incorporate additional physical ingredients into the coagulation kernel, either through interaction potentials, collision efficiency factors, or stochastic particle properties~\cite{ivlev2002coagulation-51a, suresh2022modeling, krapivsky2024,schwarzer2005prediction,ormel2007closed,millan2023monte}. In this way, the Smoluchowski framework can be extended to provide a coarse-grained description of aggregation in systems where microscopic interactions and fluctuations play an essential role.

\subsection{Structural Evolution: Compaction, Porosity, and Mechanical Strength}

Beyond their mass and size distributions, aggregates formed through coagulation processes also exhibit complex internal structures that evolve over time. Early stages of aggregation are often characterized by the formation of highly porous and loosely connected clusters, particularly when growth is dominated by low-energy collisions between small particles. Such aggregates typically exhibit fractal-like geometries in which the mass of a cluster scales sublinearly with its characteristic size~\cite{meakin1988,Dominik1997, Blum2008}. As coagulation proceeds, however, subsequent collisions may lead to significant structural rearrangements within aggregates. Rolling, sliding, and compression at contact points between constituent particles can progressively reduce porosity and increase the overall compactness of the clusters. The degree of compaction depends on factors such as collision energy, particle size, and the strength of adhesive forces at interparticle contacts. As a result, the internal morphology of aggregates can evolve from tenuous fractal structures toward more compact configurations, with important consequences for their mechanical stability, collision cross sections, and subsequent aggregation dynamics.
\begin{figure}[h!]
    \centering
    \includegraphics[width=0.8\linewidth]{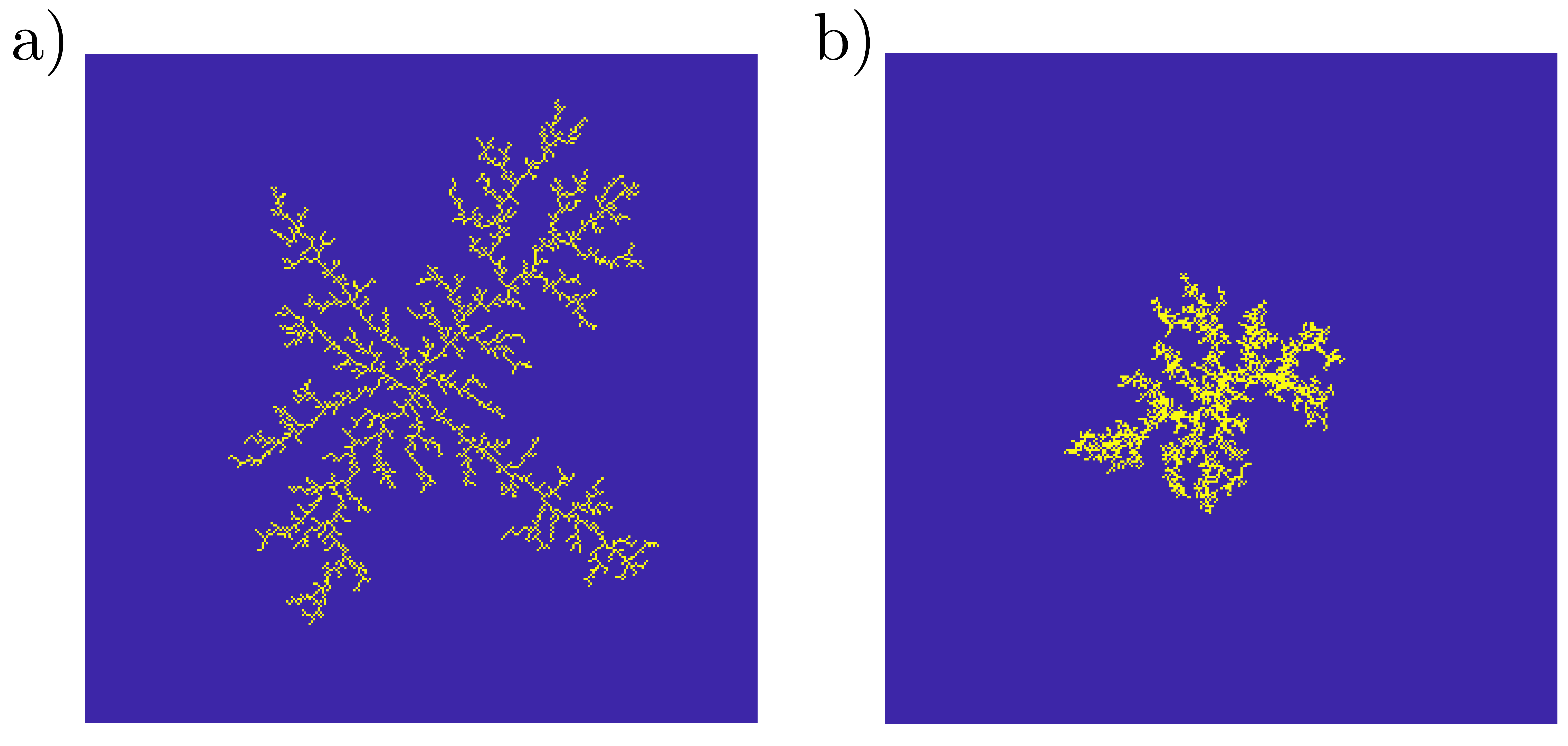}
    \caption{2D Schematic comparison of aggregate structures formed under different
aggregation regimes. In diffusion-limited cluster aggregation (DLCA),
particles stick upon first contact, producing open, fractal structures
with low fractal dimension ($D_f \approx 1.7$--$1.9$). In
reaction-limited cluster aggregation (RLCA), reduced sticking
probability allows for multiple encounters and local rearrangements,
resulting in more compact aggregates with higher fractal dimension
($D_f \approx 2.0$--$2.2$). At later stages, collisional restructuring
and compaction further increase aggregate density and mechanical
strength. These differences illustrate how microscopic collision
dynamics control the macroscopic structure and evolution of particle
aggregates. Original figure created by the authors.}
    \label{fig:DLA-RLA}
\end{figure}
A useful framework for describing the internal structure of such aggregates is provided by fractal scaling relations. In many aggregation processes, particularly those dominated by diffusion-limited or ballistic cluster--cluster collisions, the mass of an aggregate grows with its characteristic size according to a power-law relation. This behaviour is commonly expressed as $M \propto R^{D_f}$, where $M$ is the aggregate mass, $R$ is a characteristic radius (such as the radius of gyration), and $D_f$ is the fractal dimension~\cite{smirnov1990properties-2b9}. The value of $D_f$ provides a quantitative measure of aggregate compactness: low values correspond to highly porous, tenuous structures, whereas larger values indicate more compact configurations. Thus, this scaling behaviour provides a simple
framework for quantifying porosity in coagulating particle systems. Changes in the fractal dimension during collisional restructuring or compaction therefore translate directly into variations of aggregate density and mechanical stability, influencing both aerodynamic properties and the subsequent collision dynamics of the clusters.
Experimental and numerical studies have shown that different aggregation regimes lead to distinct fractal dimensions, reflecting the underlying collision dynamics and restructuring mechanisms. 
For example, in diffusion-limited cluster aggregation (DLCA), where particles stick upon contact with essentially unit probability, clusters grow through
random encounters that produce highly tenuous and open structures with
fractal dimensions typically in the range $D_f \approx 1.7$--$1.9$.
When the sticking probability is reduced, aggregation enters the
reaction-limited cluster aggregation (RLCA) regime, in which repeated
collisions are required before particles bind. This additional
rearrangement during encounters leads to more compact aggregates with
larger fractal dimensions, typically $D_f \approx 2.0$--$2.2$~\cite{weitz1984fractal, meakin1988,paszun2009collisional-d05} (Fig.~\ref{fig:DLA-RLA}). 

Collisional restructuring plays a central role in determining how the internal morphology of aggregates evolves during continued growth. When two clusters collide, the kinetic energy of the encounter may be partially dissipated through processes such as rolling, sliding, or twisting at the contact points between constituent particles. These microscopic rearrangements allow aggregates to reorganize internally, reducing void space and increasing the number of interparticle contacts~\cite{ysard2018optical-cff, YANG20233395}. As a result, repeated collisions can gradually transform highly porous fractal clusters into more compact structures~\cite{tanaka2023compression,Weidling_2009}. The efficiency of this compaction process depends on the balance between collision energy and the adhesive forces that bind particles together~\cite{2011ApJ...737...36W, 2007ApJ...661..320W}. At sufficiently low energies, aggregates may retain their initial fractal geometry with little internal rearrangement. At higher energies, however, rolling and restructuring become more likely, leading to progressive densification. Such processes are particularly relevant in granular and particulate systems where contact mechanics governs the dissipation of energy during collisions. The resulting structural evolution influences not only the porosity of aggregates but also their aerodynamic behaviour and their subsequent collision dynamics~\cite{demidov2024discrete,gunkelmann2016influence-6cb}.

The mechanical stability of particle aggregates is closely linked to their internal structure and the strength of the adhesive forces at interparticle contacts. In loosely packed aggregates composed of small grains, cohesion typically arises from surface forces such as van der Waals interactions or electrostatic attraction, which act over very short distances but can produce significant binding energies when many contacts are present~\cite{tatsuuma2019tensile}. The overall strength of an aggregate therefore depends not only on the magnitude of these microscopic forces but also on the number and spatial distribution of contacts within the structure. Assemblies with low
coordination numbers possess few load-bearing contacts and, when below
the isostatic threshold, support stress only marginally, whereas more
highly connected packings are mechanically rigid and can sustain larger
stresses before yielding \cite{Hecke2010, salinas2026unjamming-eed}. The dependence on
porosity is less direct: the compressive strength of aggregates rises far
more steeply with decreasing porosity than the tensile strength, so that
highly porous aggregates can absorb substantial energy through
compaction, whereas more compact aggregates store a larger elastic
component that may exceed the tensile strength and drive fragmentation
\cite{Tatsuuma_2023}. As compaction proceeds and additional contacts form between constituent particles, the aggregate can sustain larger stresses before breaking apart. These structural and mechanical properties play a crucial role in determining the resilience of aggregates during subsequent collisions, influencing whether encounters lead to further growth, restructuring, or fragmentation~\cite{haack2020tensile-c4a}.

Electrostatic interactions may also enhance the mechanical stability of
particle aggregates by increasing the effective binding energy between
constituent grains. In collisions between charged dielectric
agglomerates, electrostatic forces such as Coulomb attraction and
polarization can promote the re-agglomeration of fragments after
impact, thereby reducing the overall breakage ratio
\cite{Lee2015,teiser2025growth-5e6, schwaak2024high-168}. This effect is most pronounced at moderate
impact velocities, where electrostatic interactions remain comparable
to the kinetic energy of the collision. At very low velocities the
aggregate typically remains intact, while at sufficiently high
velocities the kinetic energy dominates and fragmentation occurs
despite electrostatic attraction. These results highlight how
electrostatic interactions can increase aggregate cohesion and
improve resistance to fragmentation.

Consequently, the evolution of aggregate structure and porosity is not
merely a geometric property but an important factor controlling the
long-term kinetics of coagulation processes.

\subsection{Transient Clustering at Intermediate Densities}\label{sec:saturn}

Between the dilute regime of uncorrelated binary collisions and the dense
regime of enduring force networks lies an intermediate regime, relevant
to many natural and industrial systems, where binary collisions coexist
with transient clustering. As the volume fraction increases, particles
develop spatial and velocity correlations and organise into clusters that
continually form and dissolve on dynamical timescales, progressively
violating the mean-field assumptions of the Smoluchowski equation before
the quasi-static force chains of the jammed limit develop. Saturn's rings
provide a paradigmatic illustration: local $N$-body simulations including
collisions and mutual self-gravity show that, at the moderate optical
depths of the A and B rings, particles continuously assemble into
elongated, gravitationally bound \textit{self-gravity wakes} that are
sheared apart and reform on roughly an orbital period
\cite{Salo1995,DaisakaIda1999}, their maximum size set by the competition
between self-gravity, collisional dissipation, and Keplerian shear rather
than by a sticking threshold. Cassini stellar occultations confirm such
clumping, the measured optical depth varying with viewing geometry as
expected for opaque wakes separated by nearly empty gaps
\cite{Colwell2007}. As a result, the effective kernel is both enhanced by
the increased pair correlation at contact and saturated by the continual
disruption of new aggregates, so the system reaches a dynamic steady
state rather than the runaway growth of Sec.~4a.\\
Bridging the dilute and dense descriptions here remains an open challenge.
The Smoluchowski formalism can be augmented with a disruptive term,
yielding coagulation--fragmentation balance equations whose steady states
reproduce the observed distributions; for Saturn's rings this gives a
universal power law with a large-size cutoff, in quantitative agreement
with the data \cite{Brilliantov2015}. Alternatively, the kinetic
description can be corrected for finite density through Enskog-type pair
correlations, as in the kinetic theory of dense granular gases
\cite{brilliantov2010kinetic}, or the mean-field picture can be abandoned
altogether in favour of an evolving contact network tracked in
particle-resolved simulations. A unified treatment interpolating from the
binary-collision kernel to the network-dominated limit as a function of
volume fraction remains an important goal for future work.
\subsection{From Well-Mixed Kinetics to Granular Constraints}

While the modifications discussed above account for interaction forces
and collision-level physics, they still rely on the implicit assumption
that the system remains well mixed. In many granular and particulate
systems, however, collective phenomena such as segregation, jamming,
and clogging introduce spatial and dynamical constraints that
fundamentally alter aggregation kinetics.

Granular segregation provides a clear example of how spatial
inhomogeneities can disrupt the assumptions underlying classical
coagulation theory. In ideal Smoluchowski kinetics, the system is
assumed to be well mixed, so that all particles have an equal
probability of encountering one another regardless of their size or
properties. In contrast, granular systems often exhibit size-dependent
segregation, such as the Brazil nut effect, in which larger particles
migrate toward the top of a vibrated or sheared medium while smaller
particles remain below. In dilute, fluid-coupled
systems such as aerosols, clouds, and protoplanetary disks, an analogous
separation arises instead from size-dependent gravitational settling and
drift, whereby larger particles sediment faster and become spatially
sorted from smaller ones~\cite{friedlander2000smoke}. When applied to aggregating systems, this
mechanism can lead to a spatial separation between large clusters and
small monomers, effectively reducing the collision frequency between
different size classes~\cite{rosato1987,ThorntonSegregation}. As a consequence, the hierarchical growth
process may be strongly suppressed, since large aggregates are no
longer able to efficiently capture smaller particles. In this sense,
segregation acts as a kinetic barrier that breaks the mean-field
assumption of homogeneous mixing, leading to aggregation dynamics that
are controlled by spatial organization rather than solely by collision
rates.

\begin{figure}
    \centering
    \includegraphics[width=0.8\linewidth]{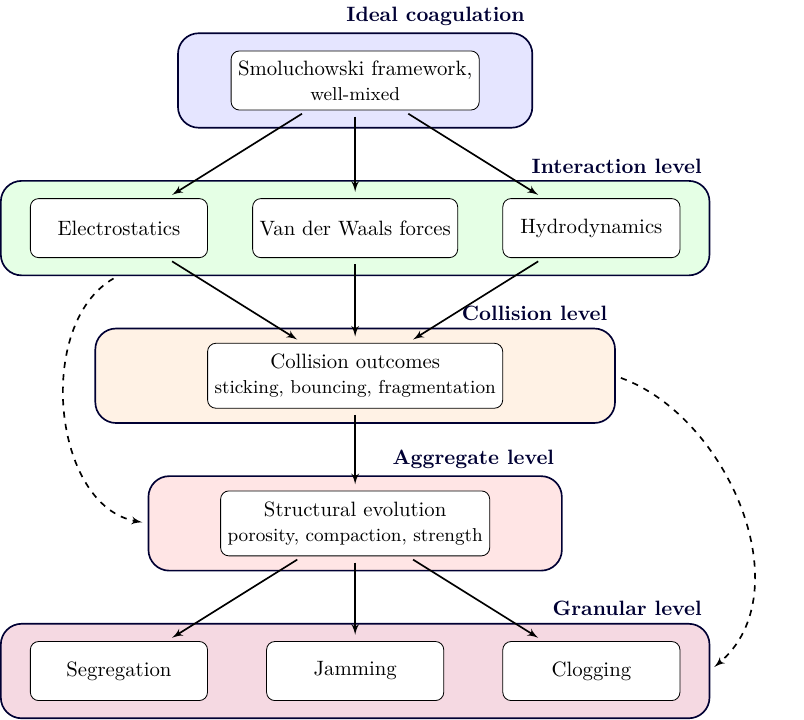}
    \caption{Schematic representation of the physical mechanisms that modify classical Smoluchowski coagulation kinetics, organized across hierarchical levels. Starting from the ideal mean-field description, which assumes a well-mixed system and binary collisions, interaction-level effects such as electrostatic forces, charge fluctuations, and other interparticle interactions alter collision probabilities. These interactions influence collision outcomes, including sticking, bouncing, and fragmentation, which in turn determine the structural evolution of aggregates through processes such as compaction and restructuring. At larger scales, collective granular effects such as segregation, jamming, and clogging introduce spatial constraints and dynamical arrest, leading to deviations from ideal kinetic behaviour. Toward the lower rows, the mean-field connections shown here progressively break down, and aggregation increasingly requires problem-specific rather than kinetic descriptions. The diagram highlights that aggregation results from the coupled interplay of mechanisms across multiple levels rather than a purely sequential process.}
    \label{fig:mechanisms}
\end{figure}

As the density of a particulate system increases, the nature of particle
interactions can shift from transient binary collisions to enduring
multi-particle contacts, leading to the formation of force-bearing
networks. This transition is commonly associated with jamming, in which
particles become mechanically constrained by their neighbours and the
system develops a finite yield stress~\cite{liu1998,song2008phase-a4b,Hecke2010}. In this state, the physics of the system is no longer described by the frequency of encounters, but by the distribution of stresses across the force network. From the perspective of
aggregation kinetics, jamming represents an extreme breakdown of the
Smoluchowski framework: the relative motion of particles is suppressed,
and the effective collision rate approaches zero. In this regime,
particles are no longer free to explore configuration space through
diffusive or ballistic motion, but are instead trapped within a
collective structure that inhibits further rearrangement. As a result,
the growth of aggregates can become effectively arrested, even in the
presence of attractive interactions that would otherwise promote
coagulation. Jamming can therefore be interpreted as a kinetic
``off-switch,'' marking a transition from a dynamically evolving system
to a mechanically stabilized state in which further aggregation is
strongly hindered.

Geometric confinement introduces an additional mechanism by which
aggregation can be limited, particularly in systems where particles
interact within restricted domains such as pores, channels, or
containers. In such environments, the growth of aggregates is not only
governed by interaction forces and collision dynamics, but also by the
available free space and the topology of the confining geometry. A
striking manifestation of this effect is clogging, where particles form
stable arch-like structures that span the confining boundaries and
block further flow~\cite{Zuriguel2014, Arevalo2015}. From the perspective of aggregation, such arches
can be interpreted as system-spanning clusters whose size is set by the
geometry rather than by intrinsic kinetic processes. Once formed, these
structures prevent additional particles from entering the region,
effectively halting further growth within the confined domain. This
highlights that, in constrained systems, aggregation kinetics can be
limited not only by energetic or dynamical considerations but also by
purely geometric constraints, which impose an upper bound on aggregate
size and alter the pathways of cluster formation. A distinct situation arises in dense or strongly sheared granular flows,
which are driven far from equilibrium and in which aggregation can
proceed only when an interparticle attraction is strong enough to survive
shear and size segregation. Examples include the magnetically induced
clusters formed in a chute flow~\cite{ThesisMagneticChuteFlow} and the
electrostatically driven aggregation of charged ash in volcanic plumes,
where Coulomb cohesion between tribocharged particles promotes
clustering and premature sedimentation within a turbulent, segregating
flow~\cite{CimarelliGenareau2022}. In the volcanic case, size segregation is moreover accompanied by charge separation: because tribocharging is size-dependent, large particles tend to charge positively and smaller ones negatively. Thus, size segregation by sedimentation also separates charges of opposite sign, generating large-scale electric fields strong enough to produce the electrical discharges observed as volcanic lightning. Such strongly out-of-equilibrium
systems typically resist a general kinetic treatment and must be addressed
case by case, in contrast to the dilute or weakly correlated regimes in
which adaptations of the Smoluchowski framework remain meaningful. In summary, as illustrated in Fig.~\ref{fig:mechanisms}, aggregation in real systems emerges from the coupled interplay of interaction,
collision, structural, and collective processes that extend beyond
the assumptions of classical mean-field kinetics.

Taken together, these effects indicate that aggregation in many
particulate systems cannot be fully described by mean-field kinetic
approaches alone. Instead, the evolution of the system reflects a
coupled interplay between particle interactions, structural
organization, and collective dynamics. In this context, aggregation
emerges not simply as a sequence of binary collisions, but as a process
constrained by the evolving spatial and mechanical state of the system.
Accounting for these effects is therefore essential for extending
coagulation theory beyond dilute systems toward realistic granular and
dense particulate environments.

\section{Applications Across Physical Systems}
\begin{figure}
    \centering
    \includegraphics[width=0.85\linewidth]{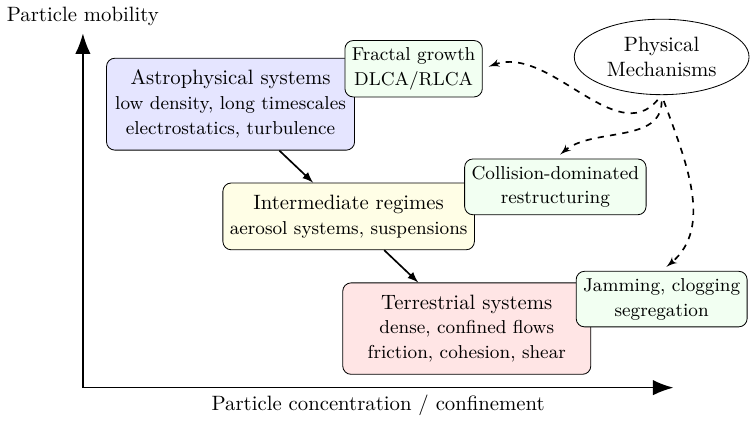}
    \caption{
Conceptual diagram of aggregation regimes across different physical systems, represented in a space defined by particle concentration (or confinement) and particle mobility (or mean free path). Dilute astrophysical environments occupy the upper-left region, where large mean free paths and low densities lead to aggregation dominated by binary collisions, long-range interactions, and fractal growth. Terrestrial granular systems lie in the lower-right region, where high densities and constrained motion give rise to contact-dominated dynamics and collective effects such as jamming, segregation, and clogging. Intermediate regimes, including aerosols and suspensions, exhibit a combination of these behaviours. Annotations indicate the dominant physical mechanisms in each region, illustrating how
aggregation transitions from collision-controlled to constraint-dominated dynamics as density increases and mobility decreases. Toward the lower-right region, the kinetic, collision-based description progressively gives way to constraint-dominated dynamics that generally require problem-specific rather than mean-field treatments.
}
    \label{fig:aggregation_regimes}
\end{figure}

The physical mechanisms discussed in the previous section illustrate
how aggregation dynamics are shaped by interactions, structure, and
collective effects that go beyond idealized kinetic descriptions.
These concepts are relevant across a wide range of physical systems,
from dilute astrophysical environments to dense terrestrial granular
media. Despite large differences in scales and conditions, many of
these systems exhibit common aggregation behaviors that can be
understood within a unified framework. In this section, we highlight
representative applications in astrophysical and terrestrial contexts,
emphasizing both shared mechanisms and system-specific features. Figure~\ref{fig:aggregation_regimes} provides a conceptual map of
aggregation regimes across physical systems, highlighting how the
dominant mechanisms vary with particle concentration and mobility.

\subsection{Astrophysical Systems: From Dust to Planetesimals}
Aggregation processes play a central role in a variety of astrophysical
environments, where the growth of solid bodies often begins from
microscopic dust grains~\cite{Birnstiel2024,Drazkowska2014,Guttler2010, wurm2021understanding-87f}. In protoplanetary disks, coagulation of
submicron particles leads to the formation of increasingly larger
aggregates, ultimately contributing to the early stages of planet
formation. Similar processes are relevant in the evolution of
planetesimals, asteroids, and cometary bodies, where collisions,
fragmentation, and restructuring govern the growth and stability of
aggregates. These systems are typically characterized by low densities,
long interaction timescales, and the importance of additional effects
such as turbulence, contact electrification, and gravitational
interactions. As a result, aggregation in astrophysical contexts
provides a rich testing ground for models that incorporate both
microscopic interactions and large-scale dynamical processes.

A central problem in astrophysics is understanding how microscopic dust
grains evolve into macroscopic bodies. The internal
structure of early aggregates is thought to influence the
properties of larger bodies, with comets in particular providing
evidence of highly porous, loosely bound material inherited from
low-energy aggregation processes~\cite{levasseurregourd2008dust-8a3,kolokolova2010comet-c0c,blum2022formation-1c0}. Asteroids, in contrast, often exhibit
more compact and mechanically processed structures, reflecting a more
complex collisional and thermal history~\cite{carry2012density-b9f,bagatin2018internal,doi:10.1073/pnas.2214353120}. At early stages, growth is
typically driven by low-velocity collisions that produce highly porous,
fractal aggregates, although additional effects such as turbulence,
electrostatic charging, and differential settling progressively modify
the aggregation dynamics~\cite{drazkowska2014can-435}.

One of the most widely discussed challenges in this context is the
presence of growth barriers associated with collision outcomes.
Laboratory experiments and numerical models have shown that, beyond a
certain size or collision velocity, particles may no longer stick upon
impact but instead bounce or fragment. This so-called bouncing barrier
and fragmentation barrier can prevent the continuous growth of
aggregates, limiting their size to millimeter or centimeter scales
under typical disk conditions~\cite{dominik2024bouncing-10e}. The competition between sticking,
restructuring, and fragmentation is therefore central to determining
whether aggregates can overcome these barriers and continue growing
toward larger bodies.

Electrostatic interactions and charge distributions may also play a
significant role in astrophysical aggregation~\cite{steinpilz2020electrical,okuzumi2011electrostatic-c05}. Charging processes, arising from plasma interactions, photoelectric effects or contact electrification, can either
suppress or enhance coagulation depending on the sign and magnitude of
particle charges. In particular, charge fluctuations and polarization
effects can lead to heterogeneous interaction strengths within the
particle population, potentially enabling the formation of larger
aggregates even in regimes where average interactions are repulsive.
These mechanisms provide possible pathways to bypass classical growth
barriers, highlighting the importance of incorporating detailed
interaction physics into models of dust evolution. Together, these
processes illustrate how aggregation in astrophysical environments
emerges from a complex interplay between kinetic, structural, and
electromagnetic effects.

\subsection{Terrestrial Granular Systems: Industrial and Geophysical Contexts}
Aggregation phenomena are also central to a wide range of terrestrial
granular systems, particularly in industrial and geophysical contexts.
In industrial processes such as fluidized beds, granulation, and powder
handling, particle aggregation and breakage determine the evolution of
size distributions and strongly influence material properties~\cite{10.1122/1.5143023}.
Unlike many astrophysical environments, these systems are often dense
and strongly influenced by frictional contacts, capillary forces, or liquid bridges, which can significantly enhance sticking probabilities~\cite{steinkogler2015granulation-1ee,Jaeger1990}. Additionally, confinement, and
collective granular effects such as segregation, jamming, and clogging~\cite{ThorntonSegregation,Zuriguel2014,liu1998} also affect the aggregation process.
These features lead to aggregation dynamics that are tightly coupled to
mechanical behaviour and flow conditions, highlighting the importance of
integrating kinetic descriptions with granular physics. At the same time, high collision energies and shear stresses may lead
to fragmentation and erosion, resulting in a dynamic balance between
growth and breakage that determines the steady-state size distribution
of particles.

As discussed in the previous
section, the strong
coupling between aggregation dynamics and collective phenomena introduce spatial heterogeneity and
dynamical constraints that are not captured by mean-field kinetic
models. For example, size segregation can lead to the spatial
separation of particle populations, reducing collision rates between
different size classes, while jamming can arrest particle motion and
effectively halt aggregation. In confined geometries, clogging and arch
formation may further limit growth by imposing geometric constraints on
particle rearrangement and transport.

Aggregation is also relevant in geophysical systems, where it influences
processes such as sediment transport, debris flows, and the formation
of aggregates in volcanic ash clouds~\cite{andreotti2013granular, BALDASSARRI2015291}. In these environments, particle
growth is often coupled to fluid flow, gravitational settling, and
cohesive interactions, leading to complex, multi-scale dynamics.
Depending on environmental conditions, aggregation may enhance
sedimentation rates, modify flow rheology, or promote the formation of
large, mechanically stable clusters. These examples highlight that, in
terrestrial systems, aggregation cannot be viewed as a purely kinetic
process but must be understood in conjunction with mechanical,
hydrodynamic, and environmental factors that govern particle motion and
interaction~\cite{andreotti2013granular}.

\section{Progress Since Blum (2006)}
In recent years, significant progress has been made in understanding
particle aggregation through a combination of experimental,
observational, and theoretical advances~\cite{Brisset2019,10.1093/mnras/stae2247,wurm2021understanding-87f}. Since the seminal work of
Blum and Wurm, new laboratory experiments and space-based observations
have provided detailed insights into collision dynamics, aggregate
structure, and growth environments, while parallel developments in
theoretical and computational modeling have enabled increasingly
realistic descriptions of aggregation processes. Together, these
approaches have revealed important deviations from idealized
coagulation models and have established aggregation as a multi-scale,
system-dependent phenomenon.
\subsection{Experimental and Observational Advances}
Laboratory experiments over the past two decades have significantly
advanced our understanding of particle aggregation by probing collision
outcomes, restructuring mechanisms, and the role of interparticle
forces under controlled conditions~\cite{brisset2016submillimetre}. In particular, a large body of work
by Blum, Wurm, and collaborators has systematically investigated
collisions between dust aggregates across a wide range of sizes and
impact velocities, establishing detailed maps of sticking, bouncing,
and fragmentation regimes~\cite{blum2000,wurm2021understanding-87f,teiser2025growth-5e6,love2014particle-a88,whizin2025particle-338,kothe2013free-b45,jungmann2021observation-d58,PhysRevLett.93.021103}. A key development has been the use of
microgravity platforms, including drop towers and parabolic flight
experiments, which enable the study of low-velocity collisions under
conditions relevant to protoplanetary disks. These experiments have
demonstrated that aggregate growth is strongly influenced by porosity,
surface properties, and internal structure, and have revealed the
existence of growth barriers such as the bouncing regime, where
particles fail to stick despite repeated collisions. More recent
experiments have also explored the role of electrostatic charging,
cohesive forces, and aggregate restructuring, highlighting the
importance of non-ideal effects in determining collision efficiency.
Together, these studies have provided essential input for the
development of physically grounded coagulation kernels and have helped
bridge the gap between microscopic collision physics and macroscopic
aggregation models.

Observational advances have provided unprecedented insight into
aggregation processes in astrophysical environments, particularly in
protoplanetary disks~\cite{andrews2020observations-a6b, Birnstiel2024,Kurtovic_2025}. High-resolution observations from facilities such
as the Atacama Large Millimeter/submillimeter Array (ALMA) have
revealed detailed substructures in disks, including rings, gaps, and
asymmetric features that are often interpreted as signatures of dust
growth and radial transport. Measurements of spectral indices and dust
emission profiles indicate the presence of millimeter- to centimeter-
sized grains, providing direct evidence of particle growth beyond the
initial stages of coagulation. These observations have challenged
simple models of continuous growth, suggesting instead that aggregation
is strongly influenced by local disk conditions, pressure traps, and
dynamical processes that concentrate particles and modify collision
rates.

Space missions have further transformed our understanding of aggregate
structure and evolution by providing in situ measurements of small
bodies and dust populations. Observations from the Rosetta mission, in
particular, have revealed that cometary material is composed of highly
porous aggregates, supporting the idea that comets preserve primordial
structures formed through low-energy coagulation~\cite{guttler2017characterization,kim2023cometary,tatsuuma2019tensile,blum2022formation-1c0}. Measurements of dust
particle properties, including size distributions, morphology, and
mechanical strength, have provided direct constraints on aggregation
models and the role of cohesive forces. Similarly, missions to
asteroids have shown that many small bodies are ``rubble piles''
consisting of gravitationally bound aggregates with complex internal
structures. A related example is the accretion of particles into meteorites. Agglomeratic olivine chondrules have been shown to contain large olivine grains ($\sim80-\SI{280}{\micro\m}$) surrounded by many smaller ones ($<1$ to $\SI{30}{\micro\m}$), which has been interpreted as evidence that these were formed via electrostatic attraction given that larger grains usually charge positively and smaller ones negatively~\cite{Schrader2018}. Together, these observations highlight the importance of
structural evolution, fragmentation, and re-aggregation processes in
shaping the properties of macroscopic bodies formed from microscopic
particles.

\subsection{Theoretical and Computational Developments}

Alongside experimental and observational progress, significant
advances have been made in the theoretical and computational
description of aggregation processes. Modern approaches extend
classical Smoluchowski theory by incorporating interaction
potentials, collision efficiency, and particle-level properties such
as charge, porosity, and morphology. In parallel, numerical methods,
including Monte Carlo simulations, discrete element models, and
population balance approaches, have enabled the study of aggregation
under increasingly realistic conditions. These developments allow for
the exploration of non-ideal effects such as fragmentation,
restructuring, and spatial heterogeneity, providing a more detailed
and flexible framework for connecting microscopic interaction physics
with macroscopic growth dynamics.

A major direction of theoretical development has been the extension of
classical coagulation kernels to incorporate additional physical
processes beyond purely geometric collision rates~\cite{PhysRevE.79.026408,okuzumi2009electric,PhysRevE.110.044128,ivlev2002coagulation-51a}. Modern formulations
often include interaction potentials, collision efficiency factors,
and particle properties such as charge, porosity, and composition,
leading to kernels that depend on multiple variables and capture
non-trivial interaction dynamics. In particular, electrostatic forces,
van der Waals interactions, and polarization effects have been
integrated into aggregation models to account for both attractive and
repulsive interactions~\cite{fuhrer2025hybrid-d52}. These extensions allow for a more accurate
description of aggregation in systems where collision outcomes are not
determined solely by encounter frequency, but by the interplay between
kinetic energy and interaction forces.

In parallel, advances in computational methods have enabled the study
of aggregation processes with increasing levels of detail. Monte Carlo
approaches based on stochastic solutions of the Smoluchowski equation
have been widely used to explore cluster size distributions under
generalized kernels, including cases with charge-dependent or
structure-dependent interactions. At the particle-resolved level,
discrete element methods and molecular dynamics simulations have been
employed to investigate collision dynamics, restructuring, and
fragmentation processes, providing direct insight into the role of
contact mechanics and energy dissipation~\cite{zhou2020discrete,wang2015monte,millan2023monte-682}. These simulations are
particularly valuable for studying regimes where analytical approaches
are intractable, such as dense systems or those exhibiting strong
heterogeneity in particle properties.

More recently, theoretical efforts have focused on going beyond
mean-field descriptions by incorporating stochastic fluctuations,
spatial correlations, and coupling to external fields or flows~\cite{brilliantov2010kinetic, liu2011kinetic-b9a, chialvo2013modified}. Models
that account for heterogeneous particle populations, including
fluctuations in charge or morphology, have revealed that rare but
highly interactive particles can significantly influence overall
aggregation dynamics~\cite{ourPRE,matthews2012charging,matthews2018discrete}. In addition, spatially resolved simulations and
continuum approaches have been developed to study the effects of
transport, confinement, and collective phenomena on coagulation.
These approaches are essential for capturing deviations from ideal
well-mixed behaviour and for connecting aggregation kinetics with the
granular and hydrodynamic processes discussed in previous sections.
Together, these developments highlight a shift toward frameworks that integrate microscopic interactions,
mesoscopic structure, and macroscopic dynamics.

\section{Open Questions and Future Directions}
Despite decades of progress, a comprehensive and predictive theory of
particle aggregation remains elusive. While classical coagulation
frameworks capture essential aspects of growth, increasing evidence
from experiments, observations, and simulations has highlighted the
importance of non-ideal effects, including interaction forces,
heterogeneity, and collective dynamics. These findings point to a
fundamental need to move beyond mean-field descriptions toward
multi-scale and system-dependent approaches. In the following, we
outline key open questions and emerging directions that are likely to
define the next stage of research in aggregation phenomena.

\subsection{Fundamental Challenges}
A central challenge in the study of aggregation processes is the
difficulty of bridging scales, from microscopic interaction physics to
macroscopic growth dynamics. While interparticle forces and collision
mechanisms determine the outcomes of individual encounters, the
emergent evolution of particle size distributions is typically
described using coarse-grained kinetic equations. Establishing a
consistent link between these levels of description remains an open
problem, particularly in systems where aggregate structure, porosity,
and internal degrees of freedom influence both collision rates and
mechanical stability. Recent efforts combining particle-resolved
simulations with population balance models have begun to address this
gap, but a fully unified framework is still lacking
\cite{PhysRevE.79.026408,liu1998,millan2023monte-682}.

A related limitation of classical approaches lies in their reliance on
mean-field assumptions, such as spatial homogeneity and uncorrelated
particle motion. In many realistic systems, however, aggregation occurs
in the presence of spatial heterogeneity, clustering, and collective
dynamics that break these assumptions. Phenomena such as preferential
concentration, confinement, and dynamical arrest introduce correlations
that can strongly modify collision rates and growth pathways,
highlighting the need for theoretical frameworks that go beyond
well-mixed kinetics. Similar deviations from mean-field behaviour have
been reported in both granular and astrophysical contexts, where
transport and collective effects play a central role
\cite{brilliantov2010kinetic,ThorntonSegregation}.

Another major challenge concerns the complexity of collision outcomes.
The transition between sticking, bouncing, erosion, and fragmentation
depends sensitively on impact velocity, particle size and structure,
and the nature of interparticle forces. Despite substantial
experimental and numerical progress, a unified description of these
processes remains elusive, and collision laws are often system-specific
and difficult to generalize. This limits the predictive power of
coagulation models, particularly in regimes where multiple collision
mechanisms coexist. Systematic experimental studies have highlighted
the existence of distinct collision regimes and growth barriers, such
as the bouncing and fragmentation thresholds
\cite{guttler2010outcome-dfe,blum2022formation-1c0}.

The role of particle variability and disorder has also emerged as a
key open question. Real systems exhibit significant heterogeneity in
properties such as size, morphology, and electric charge, leading to
broad and sometimes heavy-tailed distributions of interaction
strengths. These fluctuations can give rise to non-trivial effects,
including the disproportionate influence of rare but highly
interactive particles on aggregation dynamics. Incorporating such
stochastic variability into theoretical descriptions remains a
challenge, particularly within frameworks that rely on averaged
quantities. Recent studies have emphasized the importance of charge
fluctuations and heterogeneous particle populations in modifying
coagulation pathways
\cite{okuzumi2009electric,krapivsky2024}.

Finally, aggregation processes are often strongly coupled to their
environment, including external flows, turbulence, gravitational
effects, and geometric confinement. These factors influence particle
transport, collision rates, and structural evolution, and can give
rise to complex feedback between aggregation and system dynamics. In
dense or constrained systems, collective phenomena such as jamming,
segregation, and clogging may further limit particle mobility and
arrest growth. Developing models that consistently incorporate this
coupling between aggregation, transport, and mechanical constraints
remains an important open challenge. Such effects are particularly
relevant in both geophysical and industrial systems, where flow and
confinement strongly impact aggregation dynamics
\cite{Zuriguel2014,Trewhela_Ulloa_2024}.

Viewed through this lens, aggregation is best understood not as a
sequence of binary collisions, but as an emergent, multi-scale process
governed by the coupled physics of interactions, structure, and
collective dynamics.

\subsection{Emerging Research Directions}
A central direction of current research is the development of
models that bridge microscopic interaction
physics with macroscopic aggregation dynamics. These approaches aim to
combine particle-resolved simulations with population balance and
continuum descriptions, enabling the construction of effective kernels
that retain information about structure, porosity, and interaction
forces. Such hybrid strategies are essential for capturing the feedback
between collision-level processes and large-scale growth behaviour.
Recent efforts in aerosol and astrophysical contexts have demonstrated
the importance of coupling particle-resolved dynamics with continuum
models to capture realistic growth pathways \cite{koumoutsakos2005multiscale-044,fuhrer2025hybrid-d52}.

In parallel, data-driven and machine learning approaches are emerging
as powerful tools for modeling aggregation in complex systems. These
methods are increasingly used to infer effective collision kernels,
emulate computationally expensive simulations, and identify hidden
dependencies in high-dimensional parameter spaces. By leveraging large
datasets from simulations and experiments, data-driven models offer a
promising route to overcome the limitations of traditional analytical
descriptions. Recent studies have applied machine learning techniques
to particle systems and soft matter to predict aggregation dynamics
and structure formation with high accuracy \cite{bapst2020unveiling,han2022learning,bhattoo2023learning}.

Another important direction involves the explicit incorporation of
stochastic variability and heterogeneity into aggregation models.
Rather than focusing solely on average particle properties, recent
work emphasizes the role of distributions, fluctuations, and rare
events in shaping growth dynamics. In particular, systems with broad
or heavy-tailed distributions of interaction strengths may exhibit
qualitatively different aggregation pathways. Capturing these effects
requires stochastic and non-mean-field approaches that go beyond
deterministic kernel formulations \cite{Krapivsky_2000,leyvraz2003scaling, krapivsky2010kinetic,erban2025impact}.

The role of long-range interactions, especially electrostatic and
polarization effects, is also receiving renewed attention. Advances in
both experiments and theory have shown that charge accumulation,
fluctuations, and induced dipole interactions can significantly modify
collision rates and aggregation efficiency. Analogous behaviour arises for
magnetic interactions: dipole--dipole attraction between magnetic grains
can drive the formation of clusters and chain-like aggregates both in
sheared granular flows~\cite{Modesto2020,ThesisMagneticChuteFlow} and in
ferrofluid suspensions, where chain formation has been studied as a
canonical example of dipolar self-assembly~\cite{Gennes1969,PhysRevE.70.031504}. These effects are
particularly relevant in dilute and weakly collisional environments,
but may also influence dense systems through collective interactions.
Incorporating such long-range forces into coagulation models remains an
active area of research, particularly in systems where interaction
timescales compete with transport and collision dynamics~\cite{yoshimatsu2017selfcharging-50c,ourPRE,ivlev2002coagulation-51a}.

Finally, increasing attention is being given to aggregation in dense
and constrained systems, where collective granular effects play a
dominant role. In these regimes, particle motion is strongly influenced
by mechanical contacts, confinement, and flow conditions, leading to
phenomena such as jamming, segregation, and clogging that can limit or
arrest growth. Recent studies of particle-laden and turbulent flows
have shown that particle–flow interactions and confinement can strongly
modify transport and clustering, highlighting the need for models that
couple aggregation with granular and hydrodynamic dynamics
\cite{wang2024modeling-fe3,MARCHIOLI2025105291}.


\section{Conclusions}

In this review, we have examined the kinetics of particle aggregation
from a perspective that extends classical coagulation theory to
incorporate interaction forces, structural evolution, and collective
granular effects. While the Smoluchowski framework provides a powerful
foundation for describing aggregation in dilute and well-mixed
systems, a growing body of experimental, observational, and
theoretical work demonstrates that real particulate systems often
exhibit significant deviations from these idealized assumptions. These
deviations arise from a range of mechanisms, including electrostatic
interactions, variability in particle properties, complex collision
outcomes, and the emergence of spatial heterogeneity and dynamical
constraints in dense or confined environments.

A central theme of this work is that aggregation cannot be understood
solely as a sequence of binary collisions governed by effective
kernels, but must instead be viewed as a multi-scale process shaped by
the interplay between interactions, structure, and collective
dynamics. In particular, the incorporation of granular phenomena such
as segregation, jamming, and clogging highlights the importance of
mechanical constraints and spatial organization in limiting or
redirecting growth. At the same time, advances in experimental,
observational, and computational techniques are providing new insights
into the microphysics of collisions and the evolution of aggregates
across a wide range of systems, from protoplanetary disks to
industrial and geophysical granular flows.

Looking ahead, progress in the field will depend on the development of
predictive frameworks that can consistently connect microscopic
interaction physics with macroscopic aggregation behaviour across
different regimes. Achieving this goal will require integrating
stochastic variability, structural evolution, and environmental
coupling into models that move beyond idealized, well-mixed
descriptions. Advances in multi-scale modelling, data-driven methods,
and high-resolution experiments provide promising avenues toward this
objective, but significant challenges remain in reconciling these
approaches within a unified theoretical framework. Addressing these
challenges will be essential for establishing a comprehensive and
quantitative understanding of aggregation in complex particulate
systems.

\vskip6pt

\funding{ANID grant Fondecyt Regular 1221597}.

\bibliographystyle{RS} 
\bibliography{references} 

\end{document}